\documentclass[twocolumn]{aastex631}

\usepackage{graphicx}
\usepackage{amsmath}
\usepackage{mathtools}
\usepackage{booktabs}
\usepackage{soul}

\shortauthors{Anning Gao et al.}

\newcommand{\lmfp}{\ensuremath{\lambda_\mathrm{mfp}^{912}}}
\newcommand{\specnotation}[2]{\ensuremath{\rm #1 \, {\scriptstyle #2}}}

\defcitealias{2002ApJ...565..773T}{T02}
\defcitealias{2015MNRAS.449.4204L}{L15}
\defcitealias{2009ApJ...705L.113P}{P09}
\defcitealias{2014MNRAS.445.1745W}{W14}

\begin{document}

\title{Measuring the Mean Free Path of \specnotation{H}{I} Ionizing Photons at $3.2\leq z\leq4.6$ with DESI Y1 Quasars}

\author[0009-0000-7532-5473]{Anning Gao}
\affiliation{Department of Astronomy, Tsinghua University, Beijing 100084, China}

\correspondingauthor{Anning Gao}
\email{anninggao211@gmail.com}

\author[0000-0002-7738-6875]{Jason X. Prochaska}
\affiliation{Department of Astronomy and Astrophysics, UCO/Lick Observatory, University of California, 1156 High Street, Santa Cruz, CA 95064, USA}
\affiliation{Kavli Institute for the Physics and Mathematics of the Universe (Kavli IPMU), 5-1-5 Kashiwanoha, Kashiwa 277-8583, Japan}
\affiliation{Division of Science, National Astronomical Observatory of Japan, 2-21-1 Osawa, Mitaka, Tokyo 181-8588, Japan}

\author[0000-0001-8467-6478]{Zheng Cai}
\affiliation{Department of Astronomy, Tsinghua University, Beijing 100084, China}

\author[0000-0002-3983-6484]{Siwei Zou}
\affiliation{Chinese Academy of Sciences South America Center for Astronomy, National Astronomical Observatories, CAS, Beijing 100101, China}
\affiliation{Department of Astronomy, Tsinghua University, Beijing 100084, China}

\author[0000-0002-1991-7295]{Cheng Zhao}
\affiliation{Department of Astronomy, Tsinghua University, Beijing 100084, China}

\author[0000-0002-8246-7792]{Zechang Sun}
\affiliation{Department of Astronomy, Tsinghua University, Beijing 100084, China}

\author[0000-0001-6098-7247]{S.~Ahlen}
\affiliation{Physics Dept., Boston University, 590 Commonwealth Avenue, Boston, MA 02215, USA}

\author[0000-0001-9712-0006]{D.~Bianchi}
\affiliation{Dipartimento di Fisica ``Aldo Pontremoli'', Universit\`a degli Studi di Milano, Via Celoria 16, I-20133 Milano, Italy}

\author{D.~Brooks}
\affiliation{Department of Physics \& Astronomy, University College London, Gower Street, London, WC1E 6BT, UK}

\author{T.~Claybaugh}
\affiliation{Lawrence Berkeley National Laboratory, 1 Cyclotron Road, Berkeley, CA 94720, USA}

\author[0000-0002-1769-1640]{A.~de la Macorra}
\affiliation{Instituto de F\'{\i}sica, Universidad Nacional Aut\'{o}noma de M\'{e}xico, Cd. de M\'{e}xico C.P. 04510, M\'{e}xico}

\author[0000-0002-4928-4003]{Arjun~Dey}
\affiliation{NSF NOIRLab, 950 N. Cherry Ave., Tucson, AZ 85719, USA}

\author{P.~Doel}
\affiliation{Department of Physics \& Astronomy, University College London, Gower Street, London, WC1E 6BT, UK}

\author[0000-0002-2890-3725]{J.~E.~Forero-Romero}
\affiliation{Departamento de F\'isica, Universidad de los Andes, Cra. 1 No. 18A-10, Edificio Ip, CP 111711, Bogot\'a, Colombia}
\affiliation{Observatorio Astron\'omico, Universidad de los Andes, Cra. 1 No. 18A-10, Edificio H, CP 111711 Bogot\'a, Colombia}

\author{E.~Gaztañaga}
\affiliation{Institute of Cosmology and Gravitation, University of Portsmouth, Dennis Sciama Building, Portsmouth, PO1 3FX, UK}
\affiliation{Institut d'Estudis Espacials de Catalunya (IEEC), 08034 Barcelona, Spain}

\author[0000-0003-3142-233X]{S.~Gontcho A Gontcho}
\affiliation{Lawrence Berkeley National Laboratory, 1 Cyclotron Road, Berkeley, CA 94720, USA}

\author{G.~Gutierrez}
\affiliation{Fermi National Accelerator Laboratory, PO Box 500, Batavia, IL 60510, USA}

\author{K.~Honscheid}
\affiliation{Department of Physics, The Ohio State University, 191 West Woodruff Avenue, Columbus, OH 43210, USA}
\affiliation{Center for Cosmology and AstroParticle Physics, The Ohio State University, 191 West Woodruff Avenue, Columbus, OH 43210, USA}

\author{S.~Juneau}
\affiliation{NSF NOIRLab, 950 N. Cherry Ave., Tucson, AZ 85719, USA}

\author[0000-0001-6356-7424]{A.~Kremin}
\affiliation{Lawrence Berkeley National Laboratory, 1 Cyclotron Road, Berkeley, CA 94720, USA}

\author[0000-0002-4279-4182]{P.~Martini}
\affiliation{Department of Astronomy, The Ohio State University, 4055 McPherson Laboratory, 140 W 18th Avenue, Columbus, OH 43210, USA}
\affiliation{Center for Cosmology and AstroParticle Physics, The Ohio State University, 191 West Woodruff Avenue, Columbus, OH 43210, USA}

\author[0000-0002-1125-7384]{A.~Meisner}
\affiliation{NSF NOIRLab, 950 N. Cherry Ave., Tucson, AZ 85719, USA}

\author{R.~Miquel}
\affiliation{Institut de F\'{i}sica d’Altes Energies (IFAE), The Barcelona Institute of Science and Technology, Campus UAB, 08193 Bellaterra Barcelona, Spain}
\affiliation{Instituci\'{o} Catalana de Recerca i Estudis Avan\c{c}ats, Passeig de Llu\'{\i}s Companys, 23, 08010 Barcelona, Spain}

\author[0000-0002-2733-4559]{J.~Moustakas}
\affiliation{Department of Physics and Astronomy, Siena College, 515 Loudon Road, Loudonville, NY 12211, USA}

\author{A.~Muñoz-Gutiérrez}
\affiliation{Instituto de F\'{\i}sica, Universidad Nacional Aut\'{o}noma de M\'{e}xico, Cd. de M\'{e}xico C.P. 04510, M\'{e}xico}

\author[0000-0001-8684-2222]{J.~ A.~Newman}
\affiliation{Department of Physics \& Astronomy and Pittsburgh Particle Physics, Astrophysics, and Cosmology Center (PITT PACC), University of Pittsburgh, 3941 O'Hara Street, Pittsburgh, PA 15260, USA}

\author[0000-0001-6979-0125]{I.~P\'erez-R\`afols}
\affiliation{Departament de F\'isica, EEBE, Universitat Polit\`ecnica de Catalunya, c/Eduard Maristany 10, 08930 Barcelona, Spain}

\author{G.~Rossi}
\affiliation{Department of Physics and Astronomy, Sejong University, Seoul, 143-747, Korea}

\author[0000-0002-9646-8198]{E.~Sanchez}
\affiliation{Barcelona-Madrid RPG - Centro de Investigaciones Energéticas, Medioambientales y Tecnológicas}

\author{M.~Schubnell}
\affiliation{Department of Physics, University of Michigan, Ann Arbor, MI 48109, USA}

\author{D.~Sprayberry}
\affiliation{NSF NOIRLab, 950 N. Cherry Ave., Tucson, AZ 85719, USA}

\author[0000-0003-1704-0781]{G.~Tarl\'{e}}
\affiliation{University of Michigan, Ann Arbor, MI 48109, USA}

\author{B.~A.~Weaver}
\affiliation{NSF NOIRLab, 950 N. Cherry Ave., Tucson, AZ 85719, USA}

\author[0000-0002-6684-3997]{H.~Zou}
\affiliation{National Astronomical Observatories, Chinese Academy of Sciences, A20 Datun Rd., Chaoyang District, Beijing, 100012, P.R. China}

\begin{abstract}

The mean free path of ionizing photons { in the intergalactic medium (IGM)} ($\lambda_\mathrm{mfp}^{912}$) is a crucial quantity in modelling the ionization state of IGM and the extragalactic ultraviolet background (EUVB), and is widely used in hydrodynamical simulations of galaxies and reionization. We construct the largest quasar spectrum dataset 
to date -- 12,595 $\mathrm{S/N}>3$ spectra -- 
using the Y1 observation of Dark Energy Spectroscopic Instrument (DESI) to make the most precise model-independent measurement of the mean free path at $3.2\leq z\leq 4.6$. By stacking the spectra in 17 redshift bins and modelling the Lyman continuum profile, we get a redshift evolution $\lambda_\mathrm{mfp}^{912}\propto(1+z)^{-4.27}$ at $2\leq z\leq 5$, which is much shallower than previous estimate { $\lambda_\mathrm{mfp}^{912}\propto(1+z)^{-5.4}$}. We then explore the sources of systematic bias, including the choice of intrinsic quasar continuum, the consideration of Lyman series opacity and Lyman limit opacity evolution and the definition of $\lambda_\mathrm{mfp}^{912}$. 
Combining our results with estimates of \lmfp\ at
higher redshifts, we conclude at high confidence that 
the evolution in \lmfp\ steepens at $z \approx 5$.
We interpret this inflection as the transition from
the end of HI reionization to a fully
ionized plasma which characterizes the intergalactic
medium of the past $\sim10$~billion years.

\end{abstract}

% \keywords{Classical Novae (251) --- Ultraviolet astronomy(1736) --- History of astronomy(1868) --- Interdisciplinary astronomy(804)}

\section{Introduction} \label{sec1}

The bulk of gas existing in the vast space between galaxies, commonly referred to as intergalactic medium (IGM), is highly ionized after the epoch of reionization \citep{1965ApJ...142.1633G}. {It is believed that i}onizing photons from quasars and star-forming galaxies, which generate the extragalactic ultraviolet background (EUVB), play a crucial role in ionizing the neutral hydrogen in IGM \citep{2006AJ....132..117F, 2008ApJ...688...85F, 2009ApJ...692.1476C, 2013MNRAS.436.1023B, 2015ApJ...813L...8M, 2022NatAs...6..850J}. Hydrodynamical simulations also rely on the
EUVB, as they provide an important source of heating and photoionization to the ionized
plasma (e.g. \citealt{2006MNRAS.373.1265O, 2007MNRAS.382..325B,2015ARA&A..53...51S, 2015MNRAS.446..521S, 2019MNRAS.486.2827D, 2021ApJ...912..138V}). 

To estimate EUVB, except the photon emissivity of quasars and galaxies, it is necessary to measure the \specnotation{H}{I} Lyman limit opacity, frequently described by the mean free path of ionizing photons \citep{1996ApJ...461...20H, 2012ApJ...746..125H}. Previous studies have used various methods to measure the mean free path. Observations of high-redshift quasar spectra showed the existence of Ly$\alpha$ forest, which is composed of a series of \specnotation{H}{I} absorbers at the blueward of restframe Ly$\alpha$ emission line that follows the underlying matter distribution \citep{1998ARA&A..36..267R}. The mean free path can be calculated by integrating the distribution of column densities of these absorbers and its redshift evolution $f(N_{\specnotation{H}{I}}, z)=\partial^2 n/(\partial N_{\specnotation{H}{I}} \partial z)$ (e.g. \citealt{1993ApJ...412...34M, 2009ApJ...703.1416F, 2010ApJ...718..392P, 2013ApJ...769..146R}). However, this method relies on the assumption of the placement randomness of absorbers, thus suffering from the uncertainties from line blending and absorber clustering \citep{2014MNRAS.438..476P}. 

Another method that directly measures the mean free path exploits the composite of a large sample of quasar spectra, which is proved to significantly reduce the uncertainties \citep{2009ApJ...705L.113P}. Using this method, the mean free path has been measured at $2\lesssim z\lesssim5$ with spectra from both ground-based and space-based telescopes, yielding $\lambda^{912}_\mathrm{mfp}\propto(1+z)^{-5.4}$ \citep{2009ApJ...705L.113P, 2013ApJ...765..137O, 2013ApJ...775...78F, 2014MNRAS.445.1745W}, while other measurements show a slightly  shorter $\lmfp$ at $z\sim2.2$ \citep{2018ApJ...860...41L}. Recently the mean free path has been used to probe the end of reionization by taking into account the quasar proximity effect, which suggests a much steeper evolution at $z\gtrsim5$ indicative of a late-ending reionization model \citep{2021ApJ...917L..37C, 2021MNRAS.508.1853B, 2023ApJ...955..115Z, 2023MNRAS.525.4093G, 2024ApJ...965..134D, 2024MNRAS.530.5209R}. The mean free path can also be constrained from the individual quasar sightlines, e.g. by comparing the flux ratios between the blueward and the redward of the Lyman limit \citep{2019A&A...632A..45R}, or finding Lyman limit systems (LLS) closest to the emitting quasar using the absorption of the first 6 transitions of Lyman series{, which is less robust due to the contamination of foreground absorbers} \citep{2021arXiv210812446B}.

The Dark Energy Spectroscopic Instrument (DESI; \citealt{2016arXiv161100036D, 2016arXiv161100037D}) started its main survey of the first year in May 2021 and finished in June 2022. As a Stage-IV spectroscopic survey, DESI observed approximately $1.4$ million quasars during its first year (Y1) main survey \citep{dr1}. This achievement has resulted in today's largest quasar spectroscopic catalog, and enables today's most precise measurements of dark energy and large scale structures \citep{2024arXiv240403000D, 2024arXiv240403001D, 2024arXiv240403002D}. The large sample from DESI provides a great opportunity to measure the mean free path more precisely. In this paper, we use DESI Y1 quasar spectra to measure the mean free path at $3.2\leq z \leq4.6$ employing the updated statistical methods of \citet{2009ApJ...705L.113P} (hereafter \citetalias{2009ApJ...705L.113P}). { Thanks to the an order-of-magnitude larger sample size} compared with Sloan Digital Sky Survey (SDSS) I and SDSS-II (\citealt{2010AJ....139.2360S}, used in \citetalias{2009ApJ...705L.113P}), we obtain the measurements with the uncertainty $\sim 70\%$ lower than previous work. 
%the measurements in \citetalias{2009ApJ...705L.113P}. 
Our updated values of mean free path show a smoother redshift evolution compared with \citet{2014MNRAS.445.1745W} (hereafter \citetalias{2014MNRAS.445.1745W}), and can be useful in establishing a more precise model of EUVB. 

The rest of the paper is organized as follows. In Section~\ref{sec2} we will briefly introduce the characteristics of DESI spectra and discuss how we selected the quasar samples from its Y1 dataset. In Section~\ref{sec3} we will describe our methodology in detail, including the stacking of spectra, the theoretical model and the fitting method. We will present our results in Section~\ref{sec4} and We discuss how our refined results of mean free path could impact the modelling of other aspects in astrophysics and cosmology, particularly the modelling of reionization in Section~\ref{sec5} and summarize our conclusions in Section~\ref{sec6}. Unless otherwise specified, a flat $\Lambda$CDM cosmology is adopted, with $\Omega_m=0.3$, $\Omega_\Lambda=0.7$ and $H_0=70\,\mathrm{km}\,\mathrm{s}^{-1}\,\mathrm{Mpc}^{-1}$, and all the distances are proper distances unless stated otherwise. {For convenience, we also define the unitless $h_{70}$ to be the Hubble constant in unit of $70\,\mathrm{km}\,\mathrm{s}^{-1}\,\mathrm{Mpc}^{-1}$, which is used in our display of \lmfp\ measurements.}

\section{Data} \label{sec2}

In this section, we first summarize the properties of DESI quasars and their spectra, and discuss the potential bias that could affect our result. Then we introduce our criteria for selecting our quasar sample from DESI.

\begin{figure*}[htbp]
    \centering
    \includegraphics[width=\textwidth]{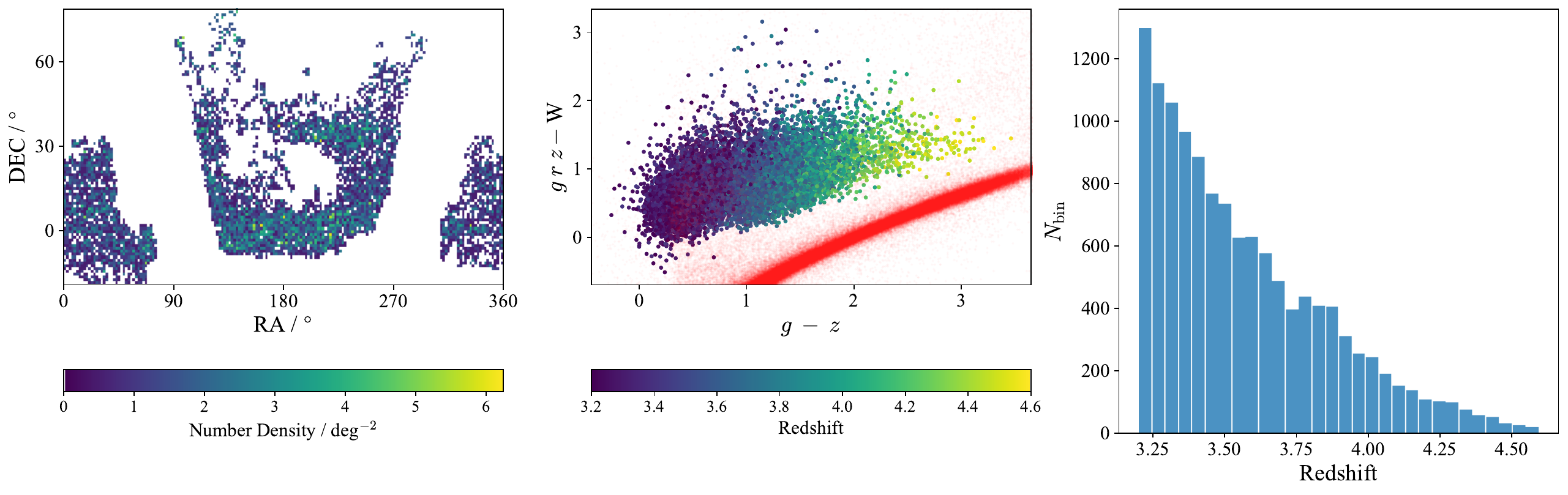}
    \caption{Parameter distributions of the DESI 
    quasar sample analyzed here. Left panel: The distribution of angular positions, showing a roughly uniform distribution across the survey area. Middle panel: The distribution of quasar magnitude and color, coloring with redshift. The $x$-axis is the $g-z$ color, and the $y$-axis is the color defined in \citet{2023ApJ...944..107C} that combines the optical $grz$ bands and the infrared W1 and W2 bands. To better illustrate the stellar contaminants, we use red points to represent the stellar color from the Early Data Realease of DESI Milky Way Survey \citep{2024MNRAS.533.1012K}, filtered with {\sc RVS\_WARN=0} and {\sc RR\_SPECTYPE=0} as well as {\sc PRIMARY=True} (see their Section 5.1). Right panel: The distribution of redshift. $N_\mathrm{bin}$ means the number of quasar sightlines within the redshift bin.}
    \label{fig1}
\end{figure*}

\subsection{The DESI Quasar Sample}

DESI was installed on the 4-meter Mayall telescope at Kitt Peak, Arizona, along with 5,000 fibers and fiber-positioners. It employs three spectrographs to obtain spectra simultaneously, commonly referred to as "b"(360-593 nm), "r"(560-772 nm) and "z"(747-980 nm), with the FWHM
resolution $\lambda/\Delta\lambda$ ranging from 2000 (blue end) to 5500 (red end). \citep{2022AJ....164..207D}. The observation is divided into dark, bright and backup programs according to the observing condition, and quasars were observed in the dark program with an effective exposure time of 
$\sim 1000$\,s \citep{2023AJ....166..259S}. CCD images are then processed to give a common 0.8\,\AA/pixel wavelength grid for all spectra \citep{2023AJ....165..144G}.

A complete and not biased quasar sample is crucial for our analysis. Here we briefly summarize the main procedure and results of DESI 
quasar target selection and validation 
\citep{2023ApJ...944..107C}. DESI incorporates the optical photometry data of $grz$ bands from the Beijing-Arizona Sky Survey, the Mayall $z$-band Legacy Survey (MzLS) and the Dark Energy Camera Legacy Survey (DECaLS) and the near infrared photometry data of W1 and W2 bands from the WISE satellite \citep{2019AJ....157..168D}, based on which a random forest algorithm is designed to select quasars in the magnitude range $16.5<r<23$. In parallel, other criteria (e.g. source morphology) were tested during the Survey Validation (SV;\citealt{2023AJ....165..124A, 2024AJ....167...62D}) to better select faint quasars and high-$z$ quasars. The final quasar catalog is constructed after a template-fitting algorithm {\sc Redrock}\citep{2024AJ....168..124A, bailey22a}, a broad \specnotation{Mg}{II} line finder and a deep convolutional neural network QuasarNet \citep{2018arXiv180809955B} are successively applied on the observed spectra, with the redshift determined at the same time. The overall procedure finally achieves a $>$99\% efficiency, a $>$98\% purity, and an approximately $60\,\mathrm{deg}^{-2}$ target density for $z>2.1$ quasars \citep{2023ApJ...944..107C}.

Despite of the excellent performance of DESI target seletion pipeline, we stress that there may still be possible biases that could affect our measurement. One of the main biases could be the quasar color selection. As discussed in \citet{2011ApJ...728...23W}, the color selection criterion of SDSS selects $3\lesssim z\lesssim 3.5$ quasars that are systematically redder on $u-g$ color and have more intervening LLSs, which is due to the difficulty of distinguishing quasars from stars in this redshift range and causes the underestimation of the mean free path measured in \citetalias{2009ApJ...705L.113P}. As we can see in \autoref{fig1}, stellar contamination still exists in DESI data at $3\lesssim z\lesssim 3.5$ and there still lacks a detailed study of DESI quasar selection bias. However, in this work we treat the DESI sample as unbiased, since the infrared observations, the more advanced random forest algorithm and the color-independent quasar variability are incorporated in DESI main selection \citep{2023ApJ...944..107C}. We still remain cautious for the potential biases, and careful examination is needed in future studies.

\subsection{Sample Selection} \label{sec22}

We select our sample of 12,595 quasars from the main survey, dark program of DESI Y1 observation. In particular, we apply the following selection rules:
\begin{enumerate}
    \item $3.2\leq z\leq4.6$, where the redshift is determined by {\sc Redrock}. This range is set by our measuring method. As we will see in Section~\ref{sec3}, our method requires the stacking of spectra and the fitting of the spectrum below the restframe Lyman limit (912\,\AA). There is only a small number of $z>4.6$ quasars that meet the other rules, which 
    was deemed insufficient for stacking. 
    Below $z<3.2$, quasars have a very short spectrum below the restframe Lyman limit which is 
    too little for robust fitting.
    \item ZWARN=0 to avoid problematic redshift determination, where ZWARN is one of the outputs of {\sc Redrock} \citep{2023AJ....165..144G, 2024AJ....168...58D, 2024AJ....167...62D}.
    \item Avoid broad absorption line (BAL) systems. Specifically, we require that the intrinsic absorption index (AI; \citealt{2002ApJS..141..267H}) and the balnicity index (BI; \citealt{1991ApJ...373...23W}) of \specnotation{C}{IV} and \specnotation{Si}{IV} are all zero, which are derived from the future Data Release 1 BAL catalog and calculated using the methods in \citet{2024MNRAS.532.3669F}.
    \item The signal-to-noise-ratio (S/N) is larger than 3, {where S/N is estimated using the average pixel-level S/N at restframe wavelength range 1450-1470\,\AA.} Following \citetalias{2009ApJ...705L.113P}, this rule is meant to decrease the variation in the stacking spectra. We also construct other samples for S/N cut of 5, 7 and 10 to perform our following fitting procedures, and our results are shown to be insensitive to the selection of S/N cut.
\end{enumerate}

After selection, the official pipeline\footnote{\url{https://github.com/desihub/desispec}} is used to combine the three cameras. The galactic extinction is corrected both for flux and error to maintain S/N, using the extinction function in \citet{1989ApJ...345..245C} ($R_V=3.1$) and the dust extinction map in \citet{1998ApJ...500..525S}. Distributions of angular location, color and redshift of our sample are summarized in \autoref{fig1}. {For comparison, we also list the measurements without this correction in Table~\ref{tab1}, which mostly show a difference less than $1\sigma$ and do not have a significant impact on the fitting. Therefore we will not use these measurements in what follows.}

\section{Methodology} \label{sec3}

In this section, we detail our technique of mean free path measurements. Following \citetalias{2014MNRAS.445.1745W}, we define the mean free path as the mean distance that the ionizing photons will travel before suffering an $1/e$ amount of attenuation, where $e$ is the base of natural logarithms.

\subsection{Stacking}

\begin{figure}
    \centering
    \includegraphics[width=\columnwidth]{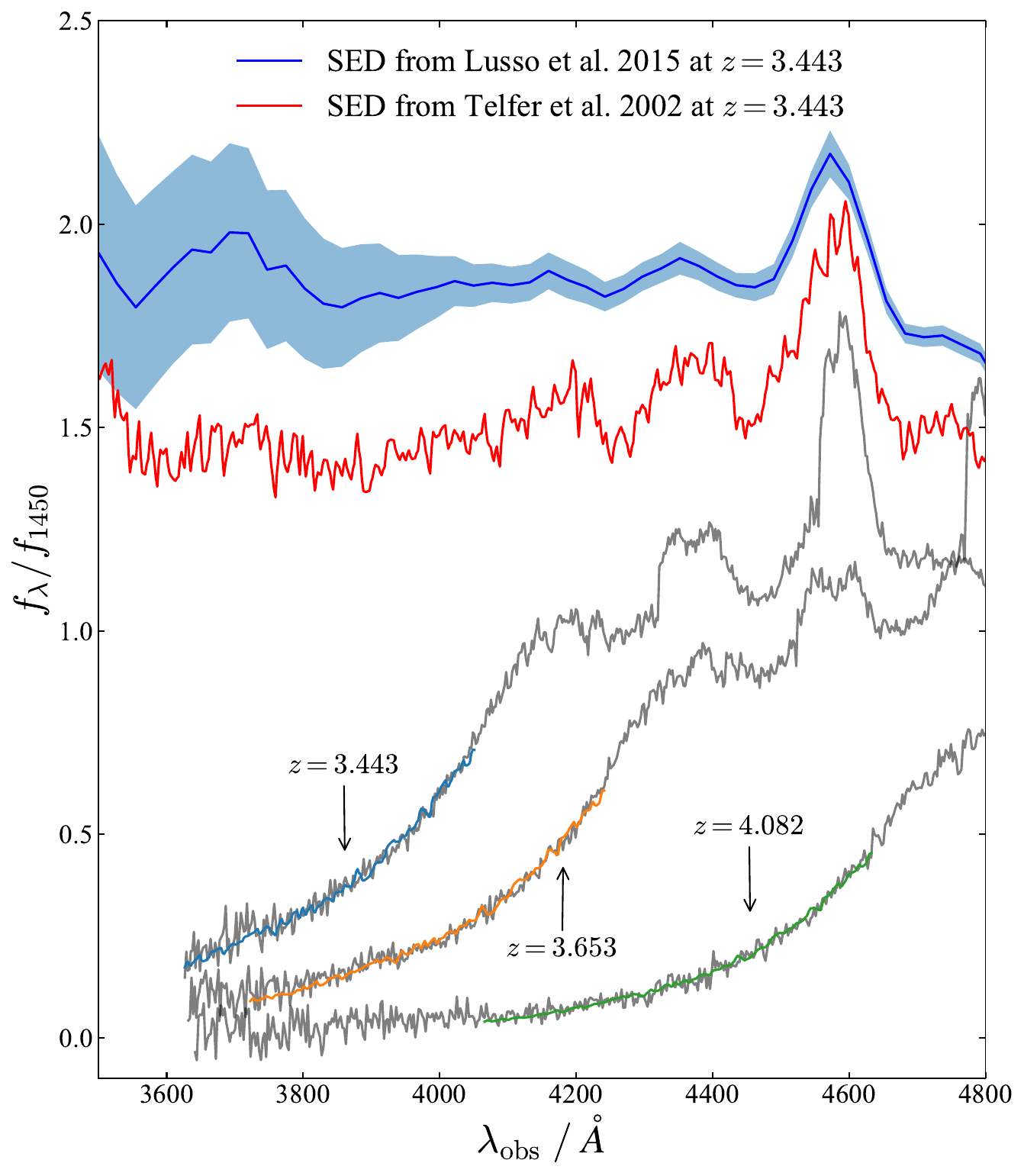}
    \caption{Examples of our stacked spectra. The grey spectra are the $z_\mathrm{median}=3.443$ (left), $z_\mathrm{median}=3.653$ (middle) and $z_\mathrm{median}=4.083$ (right) stacks, shifted to the observation frame. Overplotted are the best fitted model. The red spectrum is the \citetalias{2002ApJ...565..773T} SED, and the blue spectrum at the top is the \citetalias{2015MNRAS.449.4204L} SED, with the lighter blue region indicating the $1\sigma$ confidence region. The two SEDs are both shifted to $z_\mathrm{qso}=3.443$ to be comparable with the leftmost stack. All the five spectra are normalized at 1450\AA. Overplotted on the DESI stacked spectra are the best-fitting models using \citetalias{2002ApJ...565..773T} SED. An illustration of the composite spectra and SED templates redward of Lyman$\alpha$ is shown in Appendix~\ref{app0}.
    }
    \label{fig2}
\end{figure}

Since mean free path is a statistical quantity, the stacking of spectra is important to our analysis. Here we follow the procedures in \citet{2013ApJ...765..137O}. We divide the quasar sample into 17 redshift bins, with 1,000 or 500 quasars in each bin, except that the last one has 95 quasars. The reason why the higher redshift bins have only 500 quasars instead of 1000 is that we want to keep a balance between the redshift uncertainty and the $\lmfp$ uncertainty. The detailed information of each bin is shown in Table~\ref{tab1}. For each redshift bin, the spectra are shifted to the quasar's restframe and normalized with the median flux at 1450-1470\,\AA. Since the wavelength grid size after redshift-shifting is $0.8\text{\,\AA}/(1+z_\mathrm{qso})<0.2\text{\,\AA}$ for all the spectra, where $z_\mathrm{qso}$ is the quasar redshift, we choose a 0.5\,\AA/pixel wavelength grid for stacking, ensuring that for each pixel in the stacked spectra, there is at least one pixel from every original spectra that can be stacked into that pixel. The stacked flux is chosen to be the mean of stacked pixels. The stacked flux uncertainty is calculated as:
\begin{equation}
    \sigma_\mathrm{stacked}=\frac{1}{n}\sqrt{\sigma_1^2+\sigma_2^2+\cdots+\sigma_n^2}\label{eq1}
\end{equation}
i.e. we treat the flux as independent Gaussian distribution (For the independence, see \citealt{2023AJ....165..144G}, Section 4.5). We give every sightline an equal stacking weight, as each sightline represents a spatial direction that should be treated equally. All the pixels with a non-zero maskbit are neglected during stacking. We stress that we do not use the inverse variance weighting. Although it would of course reduce the variance of our composite spectra, it would also introduce an unwanted prior that gives high S/N sightlines more weights and may bias our result. \autoref{fig2} shows several examples of our stacked spectra.

 To estimate the sample variance and the uncertainties introduced by redshift error, we use the bootstrap technique. For each redshift bin, a random sample is generated from the quasars of this bin, allowing for duplication. Then each spectrum's redshift is allowed for a random Gaussian change, with 0 mean and the standard deviation of the redshift error equivalent to $350\;\mathrm{km/s}$, which is inspired by the visual inspection results of DESI quasars \citep{2023AJ....165..124A}. Those spectra are stacked using the procedures described before and shares the same wavelength grid with the previous stacks. This process is performed 1,000 times, and 1,000 different stacks are generated, which are then used to perform Markov Chain Monte Carlo (MCMC) analysis, as described in Section~\ref{sec33}.

\subsection{Modelling}

We use the technique in \citetalias{2009ApJ...705L.113P} to analyze the stacked spectrum at $\lambda <912$\,\AA, which is widely used in previous mean free path measurements (\citealt{2013ApJ...775...78F, 2013ApJ...765..137O}; \citetalias{2014MNRAS.445.1745W}; \citealt{2018ApJ...860...41L}). 
This portion of the spectrum is modelled to be the intrinsic spectral energy distribution (SED) attenuated by Lyman series effective optical depth, $\tau^\mathrm{LyS}_{\lambda_r}$, which is caused when the photons with restframe wavelength ${\lambda_r}$ is redshifted to a specific Lyman series line (Ly$\alpha$, Ly$\beta$, etc.), and Lyman continuum effective optical depth, $\tau^\mathrm{LyC}_{\lambda_r}$, which is caused when the photon's wavelength is shorter than 912\,\AA\;so that it can ionize \specnotation{H}{I} atoms. 
We can write down the explicit expression:
\begin{equation}
    f_{\lambda_r}=f^\mathrm{SED}_{\lambda_r}\exp{\left(-\tau^\mathrm{LyS}_{\lambda_r}\right)}\exp{\left(-\tau^\mathrm{LyC}_{\lambda_r}\right)} \label{eq4}
\end{equation}
where $f_{\lambda_r}$ is the observed composite spectrum and $f^\mathrm{SED}_{\lambda_r}$ is the intrinsic quasar SED. The three parts of \autoref{eq4} are modelled as follows, following \citet{2013ApJ...765..137O}, \citetalias{2014MNRAS.445.1745W} and \citet{2018ApJ...860...41L}:
(i) The quasar SED is assumed to be a template modified by a power law:
\begin{equation}
    f^\mathrm{SED}_{\lambda_r}=f_0f^\mathrm{tem}_{\lambda_r}\left(\frac{\lambda_r}{912\text{\,\AA}}\right)^{-\alpha} \label{eq5}
\end{equation}
where {$\alpha$ is the tilt index,} $f_0$ is a normalization factor accounting for the continuum uncertainty and $f^\mathrm{tem}_{\lambda_r}$ is the SED template. {We do not include the sky zero point as a free parameter in our above formalism, because our fitted $f_0$ is very close to 1 in all our fittings. Therefore, the effect of sky zero point uncertainty is accounted for in $f_0$ to a first-order approximation.}  \citet{2002ApJ...565..773T} (hereafter \citetalias{2002ApJ...565..773T}) and \citet{2015MNRAS.449.4204L} (hereafter \citetalias{2015MNRAS.449.4204L}) both stacked quasar spectra from \textit{Hubble Space Telescope} (\textit{HST}) and obtained average SEDs differing 
by $\sim 30$\% at Lyman continuum. For comparison, we use the SED template from both two composites and show the results in \autoref{tab1}. The base of the power law modification is chosen to be $912$\AA\; instead of the widely used $1450$\AA\; because we only want this modification to impact the slope of SED at Lyman continuum, and $1450$\AA\;base would introduce an additional overall normalization at Lyman continuum that could be degenerate with $f_0$. 

(ii) The Lyman series optical depth is described by a power law:
\begin{equation}
    \tau^\mathrm{LyS}_{\lambda_r}=\tau^\mathrm{LyS}_0\left(\frac{1+z_{912}}{1+z_\mathrm{qso}}\right)^{\gamma_\tau}\label{eq6}
\end{equation}
where $z_\mathrm{qso}$ is the quasar redshift and is represented by the median redshift of stacked spectra{ (the mean redshift is similar, as shown in Table~\ref{tab1}, so the freedom of choosing median or mean would not introduce any inconsistency)}, and $z_{912}$ is defined as:
\begin{equation}
    1+z_{912}=\frac{\lambda_r}{912\text{\,\AA}}(1+z_\mathrm{qso})
\end{equation}
i.e. $z_{912}$ represents the redshift at which the $\lambda_r$ photon is shifted to 912\,\AA. This power law is inspired by \citet{2014MNRAS.438..476P} and the index $\gamma_\tau$ is chosen to be 3 for all redshift bins, which are justified in Appendix \ref{app2}. In practice, the coefficient $\tau_0^\mathrm{LyS}$ is determined by the absorption at restframe 912\,\AA. We stress that this term is the key difference between our measurements and the analysis in \citetalias{2009ApJ...705L.113P}, where the authors only considered the Lyman continuum opacity and measured a mean free path systematically higher than ours. 
A detailed explanation is given in Section~\ref{sec4}.

(iii) The Lyman limit optical depth is described by (\citetalias{2009ApJ...705L.113P}; \citealt{2013ApJ...765..137O}):
\begin{equation}
    \tau^\mathrm{LyC}_{\lambda_r}=\int\kappa\,\mathrm{d}r
\end{equation}
where $\kappa$ is the opacity coefficient, whose redshift evolution and frequency dependence is modelled separately:
\begin{align}
\begin{split}
     \kappa=&\kappa_{0}\left(\frac{1+z}{1+z_\mathrm{qso}}\right)^{\gamma_\kappa}\left(\frac{\nu}{\nu_{912}}\right)^{-2.75}\label{eq9}\\
    =&\kappa_0\left(\frac{1+z}{1+z_\mathrm{qso}}\right)^{\gamma_\kappa}\left(\frac{1+z}{1+z_{912}}\right)^{-2.75}
\end{split}
\end{align}
{(where $\nu_\mathrm{912}$ is the Lyman limit frequency)} and $\mathrm{d}r$ is modelled with cosmology:
\begin{equation}
    \frac{\mathrm{d}r}{\mathrm{d}z}=\frac{c}{H_0}\frac{1}{(1+z)\sqrt{\Omega_m(1+z)^3+\Omega_\Lambda}}\approx\frac{c}{H_0\sqrt{\Omega_m}}(1+z)^{-2.5}
\end{equation}
The universe is matter-dominated at $3<z<5$, which justifies the last approximation in the above equation. Combining the three equations, we get:
\begin{equation}
\tau^\mathrm{LyC}_{\lambda_r}=\frac{c\kappa_0}{H_0\sqrt{\Omega_m}}\frac{(1+z_{912})^{2.75}}{(1+z_\mathrm{qso})^{\gamma_\kappa}}\int_{z_{912}}^{z_\mathrm{qso}} (1+z)^{\gamma_\kappa-5.25}\,\mathrm{d}z\label{eq12}
\end{equation}

In conclusion, there are three parameters in our model that can be fitted: the normalization factor $f_0$, the "effective" opacity coefficient $\kappa_{912}\equiv c\kappa_0/(H_0\sqrt{\Omega_m})$ and the evolution index $\gamma_\kappa$. To simplify our model, we assume $\gamma_\kappa=\alpha=0$. This comes for two reasons: First, in our redshift range, the opacity does not evolve rapidly; Second, for the fitted wavelength range the spectrum's decline is dominated by opacity, and unless there is an extreme tilt (e.g. $|\alpha| >0.5$), the result will not be highly influenced by it. 

Finally, our definition of $\lambda_\mathrm{mfp}^{912}$ corresponds to a photon that has a specific emission wavelength $\lambda_r$ and experiences a total Lyman limit optical depth $\tau^\mathrm{LyC}_{\lambda_r}=1$, from $z_\mathrm{qso}$, where the photon is emitted, to $z_{912}$, where the photon is redshifted enough to have $h\nu < 1$\,Ryd. Thereby, $\lambda_\mathrm{mfp}^{912}$ is defined to be the physical distance between $z_\mathrm{qso}$ and $z_\mathrm{912}$ for these photons:
\begin{equation}
    \lmfp\equiv\int c\,\mathrm{d}t=\int_{z_{912}}^{z_\mathrm{qso}}\frac{c\,\mathrm{d}z}{H(z)(1+z)}\label{eq10}
\end{equation}

\subsection{Fitting}\label{sec33}

We use the MCMC technique to estimate the posterior distribution of parameters for the composite spectra in each redshift bin. To better combine the uncertainty introduced by spectrum fluctuations, sample selection and redshift estimation, we perform MCMC analysis for each bootstrap spectrum, and combine all the posterior with equal weights to give the final posterior. We choose the traditional form of $\chi^2$ likelihood:
\begin{equation}
    \chi^2=\sum_{i}\left(\frac{f_i-f_i^\mathrm{model}}{\sigma_i}\right)^2\label{eq11}
\end{equation} 
where $f_i$ and $f_i^\mathrm{model}$ are the stacked flux and the model flux at the $i$th pixel and the uncertainty $\sigma_i$ is defined in \autoref{eq1} and is chosen to be the same for all the bootstrap spectra in each redshift bin to reduce the computation cost. We do not choose to represent the sample variance with the covariance matrix calculated from the bootstrap spectra, as we discovered that this would underestimate the final uncertainties of $\lmfp$, which may be probably caused by the ignorance of non-Gaussian variance. In our fiducial model, the prior is chosen to be a uniform distribution with a loose boundary $0<f_0<3$, $0<\kappa_{912}<1000$. The lower bound of the fitted spectral region 
is the maximum between 800\,\AA\;and the shortest wavelength that has a
contribution from all of the stacked spectra.

An example of combined posterior distributions is shown in {in Appendix~\ref{appMC}}, together with the mean free path distribution calculated from 200,000 sample points. Although the combined posterior is not strictly Gaussian, the marginal distributions of parameters and the mean free path distribution are still symmetric and can be viewed as an imperfect Gaussian shape. Therefore, the fitted value of the mean free path is chosen to be the median, and the uncertainty is chosen to be the standard deviation, which corresponds to the 68\% confidence level. 

\section{Results} \label{sec4}

Following the method, we fit the mean free path of 17 redshift bins using the stacked spectra from 12,595 quasars. All the fitted values are shown in \autoref{tab1}, {where only the statistical uncertainties given by MCMC are included in the error bars.} In Section~\ref{sec41} we explore the redshift evolution of the mean free path and discuss how our results differ from previous measurements that are listed in Appendix~\ref{app1}. In Section~\ref{sec42} we validate our results and discuss the possible systematic uncertainties in our measurements.

\begin{table*}
  \centering
  \caption{$\lambda^{912}_\mathrm{mfp}$ measurements. The unit of $\lambda^{912}_\mathrm{mfp}$ and $A$ is $h_{70}^{-1}\,\mathrm{Mpc}$. {Values in the brackets are measured without correction for Galactic extinction. Parameters $A$ and $\eta$ are fitted using median redshifts.}}
    \begin{tabular}{cccccccc}
    \toprule
    $z_\mathrm{min}$  & $z_\mathrm{max}$  & $z_\mathrm{median}$ & {$z_\mathrm{mean}$} & $N_\mathrm{qso}$   & {$\lambda_r$ range} & $\lambda^{912}_\mathrm{mfp}$ with \citetalias{2002ApJ...565..773T} SED   & $\lambda^{912}_\mathrm{mfp}$ with \citetalias{2015MNRAS.449.4204L} SED  \\
    \midrule
    3.200 & 3.236 & 3.217 & {3.218} & 1000  & 857-912\,\AA & 65.4{(64.7)}$\pm$3.44  & 59.7$\pm$2.83  \\
    3.236 & 3.275 & 3.255 & {3.255} & 1000  & 850-912\,\AA & 64.6{(62.7)}$\pm$2.87  & 59.1$\pm$2.38  \\
    3.275 & 3.320  & 3.297 & {3.297} & 1000  & 842-912\,\AA & 66.5{(65.2)}$\pm$3.01  & 60.5$\pm$2.49 \\
    3.320  & 3.364 & 3.341 & {3.342} & 1000  & 834-912\,\AA & 60.1{(59.0)}$\pm$2.58  & 55.1$\pm$2.16 \\
    3.364 & 3.415 & 3.389 & {3.389} & 1000  & 825-912\,\AA & 63.3{(61.4)}$\pm$2.72  & 57.6$\pm$2.24 \\
    3.415 & 3.474 & 3.443 & {3.444} & 1000  & 816-912\,\AA & 60.6{(59.3)}$\pm$2.41 & 55.3$\pm$2.00  \\
    3.474 & 3.539 & 3.506 & {3.506} & 1000  & 805-912\,\AA & 52.0{(50.9)}$\pm$1.96  & 48.0$\pm$1.68 \\
    3.539 & 3.613 & 3.577 & {3.576} & 1000  & 800-912\,\AA & 50.4{(49.2)}$\pm$1.77  & 46.6$\pm$1.51 \\
    3.613 & 3.696 & 3.653 & {3.653} & 1000  & 800-912\,\AA & 47.1{(46.4)}$\pm$1.64  & 43.7$\pm$1.42 \\
    3.696 & 3.756 & 3.724 & {3.725} & 500   & 800-912\,\AA & 43.7{(43.1)}$\pm$2.06  & 40.7$\pm$1.78  \\
    3.756 & 3.808 & 3.782 & {3.783} & 500   & 800-912\,\AA & 41.4{(40.9)}$\pm$1.97   & 38.6$\pm$1.72 \\
    3.808 & 3.866 & 3.837 & {3.837} & 500   & 800-912\,\AA & 41.3{(40.6)}$\pm$2.11  & 38.5$\pm$1.85  \\
    3.866 & 3.931 & 3.896 & {3.897} & 500   & 800-912\,\AA & 35.4{(34.7)}$\pm$1.73  & 33.3$\pm$1.53  \\
    3.931 & 4.024 & 3.976 & {3.976} & 500   & 800-912\,\AA & 35.9{(35.2)}$\pm$1.52  & 33.7$\pm$1.35  \\
    4.024 & 4.161 & 4.082 & {4.085} & 500   & 800-912\,\AA & 31.6{(31.1)}$\pm$1.38 & 29.8$\pm$1.23   \\
    4.161 & 4.434 & 4.270 & {4.278} & 500   & 800-912\,\AA & 29.5{(28.8)}$\pm$1.15 & 27.8$\pm$1.03   \\
    4.434 & 4.600 & 4.489 & {4.498} & 95    & 800-912\,\AA & 23.5{(23.5)}$\pm$1.90  & 22.4$\pm$1.72  \\
    \midrule
    $A$ &       &       & &       &  & $35.03{(34.33)}\pm0.54$ & $32.92\pm0.47$ \\
    $\eta$ &       &     & &       &  & $-4.03{(-4.06)}\pm0.16$ & $-3.84\pm0.14$ \\
    \bottomrule
    \end{tabular}
  \label{tab1}
\end{table*}

\begin{figure*}[htbp]
    \centering
    \includegraphics[width=\textwidth]{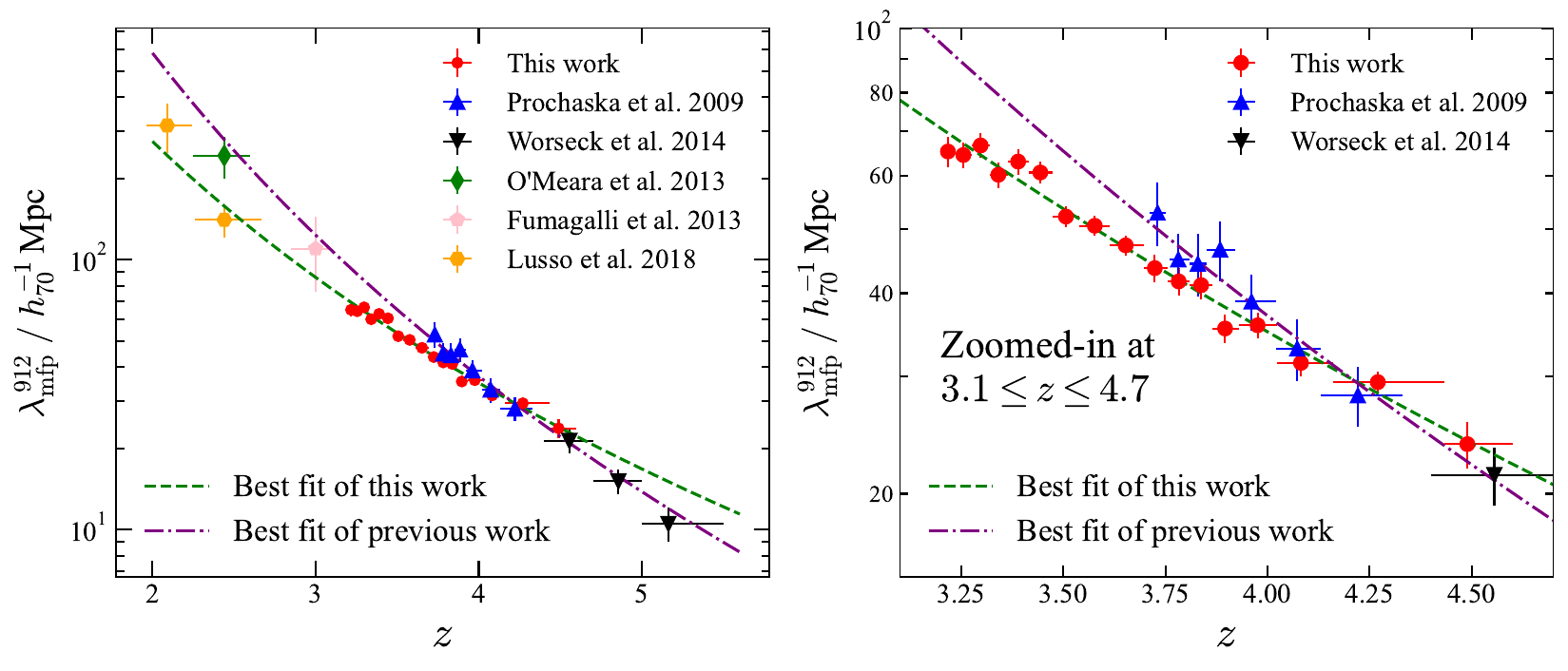}
    \caption{The measured mean free path as a function of redshift, using \citetalias{2002ApJ...565..773T} SED. Left panel: The overall mean free path evolution at $2\lesssim z\lesssim 5$ with data points from \citetalias{2009ApJ...705L.113P} (blue triangle-ups), \citet{2013ApJ...775...78F} (pink pengatons), \citet{2013ApJ...765..137O} (green diamonds), \citetalias{2014MNRAS.445.1745W} (black triangle-downs), \citet{2018ApJ...860...41L} (orange hexagons) and our work (red circles). The violet dash-dot line and the green dash line represent the best power law fitting in \citetalias{2014MNRAS.445.1745W} and our work, respectively. Right panel: The zoom-in plot at $3.1\leq z\leq4.7$ to better display the differences between our work and \citetalias{2009ApJ...705L.113P}. Our measurements are systematically lower than the interpolation of \citetalias{2014MNRAS.445.1745W} power law at $z\lesssim4$, which is mainly due to the improper handling of Lyman series opacity evolution in \citetalias{2009ApJ...705L.113P},  and our power law is roughly consistent with other previous works.}
    \label{fig5}
\end{figure*}

\subsection{Redshift Evolution of the Mean Free Path} \label{sec41}

The redshift evolution of the mean free path plays a crucial role in understanding the evolution of the thermal state of IGM and determining the content of cosmic residual neutral hydrogen at post-reionization epoch. \citetalias{2014MNRAS.445.1745W} provides a comprehensive collection of $\lambda^{912}_\mathrm{mfp}$ and gives $\lambda^{912}_\mathrm{mfp}\propto(1+z)^{-5.4\pm0.4}$, which is much steeper than the speed caused by cosmic expansion only and indicates the decrease of the number and/or size of absorbers. {This evolution is also consistent with hydrodynamical simulations in \citet{2020ApJ...898..149D}, where the IGM is fully ionized and relaxed.} \citet{2018ApJ...860...41L}, however, suggests a slower evolution of the mean free path, $\lambda^{912}_\mathrm{mfp}\propto(1+z)^{-4.5\pm0.2}$, incorporating the new measurements with an updated quasar SED. 

In recent years the mean free path measurements have been extended to higher redshifts, where the mean free path becomes too short to neglect the proximity effect of quasars. This effect is caused by the extra ionization of neutral hydrogen from the nearby quasars, and may bias the mean free path measurements as it is the local effects that dominates and 
do not reflect the opacity of the entire IGM. \citet{2021MNRAS.508.1853B} and \citet{2023ApJ...955..115Z} measures the mean free path at $5\lesssim z\lesssim6$ by considering the size of quasar proximity zone, and their measurements implies a much steeper evolution of mean free path at this redshift range.

Here we present our measurements of the mean free path using \citetalias{2002ApJ...565..773T} SED as a function of redshift in \autoref{fig5}, together with previous measurements at $z\lesssim 5$. The \citetalias{2002ApJ...565..773T} results are used as the fiducial results throughout the rest of this paper, as most previous measurements at similar redshifts, e.g. \citetalias{2009ApJ...705L.113P}, \citet{2013ApJ...765..137O}, \citet{2013ApJ...775...78F} and \citetalias{2014MNRAS.445.1745W} all used the \citetalias{2002ApJ...565..773T} SED, and employing the same SED would help us maintain consistency and compare with these results. We also fit the two-parameter model in \citetalias{2014MNRAS.445.1745W}, $\lambda^{912}_\mathrm{mfp}=A\left[(1+z)/5\right]^\eta$ using MCMC and our data \textit{only} and get $A=35.03\pm0.54\,h_{70}^{-1}\,\mathrm{Mpc}$ and $\eta=-4.03\pm0.16$. The two parameters fitted with our other measurements are shown at the bottom lines in \autoref{tab1}.

Our measurements show several major differences comparing with the previous: 
\begin{enumerate}
    \item The uncertainties are significantly reduced due to our large sample. However, as shown in Table~\ref{tab1}, although we use 1,000 spectra for stacking for the first 9 redshift bins and less spectra for the others, the $\lmfp$ uncertainties seem not to show corresponding smaller values for the first 9 bins, which may put a question on the necessity of employing such a large sample. We admit that it is indeed not clear if we look at the \textit{absolute} value, but this is not the case if we look at the \textit{relative} uncertainties. As shown in \autoref{fig7}, the first 9 bins do exhibit smaller relative uncertainties, especially if we compare the 8th and 9th points with the 10th and 11th points. The seemingly large relative uncertainties for the first 7 bins are mainly due to the shorter restframe wavelength range that we use to fit, since a shorter wavelength range gives less spectrum points available for our fitting and thus limits the constraining power.
    \item The measured mean free path are systematically lower than previous estimates, especially \citetalias{2009ApJ...705L.113P}. The discrepancy occurs mostly at $z\lesssim4$, and the four points with lowest redshift in our measurements are only $\sim 75\%$ the corresponding values of the \citetalias{2014MNRAS.445.1745W} power law. The discrepancy implies a systematic bias induced from either the data collection, i.e. the different target selection of DESI and SDSS, or the different fitting procedure. We explore a number of possibilities in the next sub-section.
\end{enumerate}

\subsection{Validations on Systematics} \label{sec42}

To further investigate the sources of 
difference in \autoref{fig5}, we conduct various validations on our fitting method. 

\begin{figure*}[htbp]
    \centering
    \begin{minipage}{\columnwidth}
        \centering
        \includegraphics[width=\columnwidth]{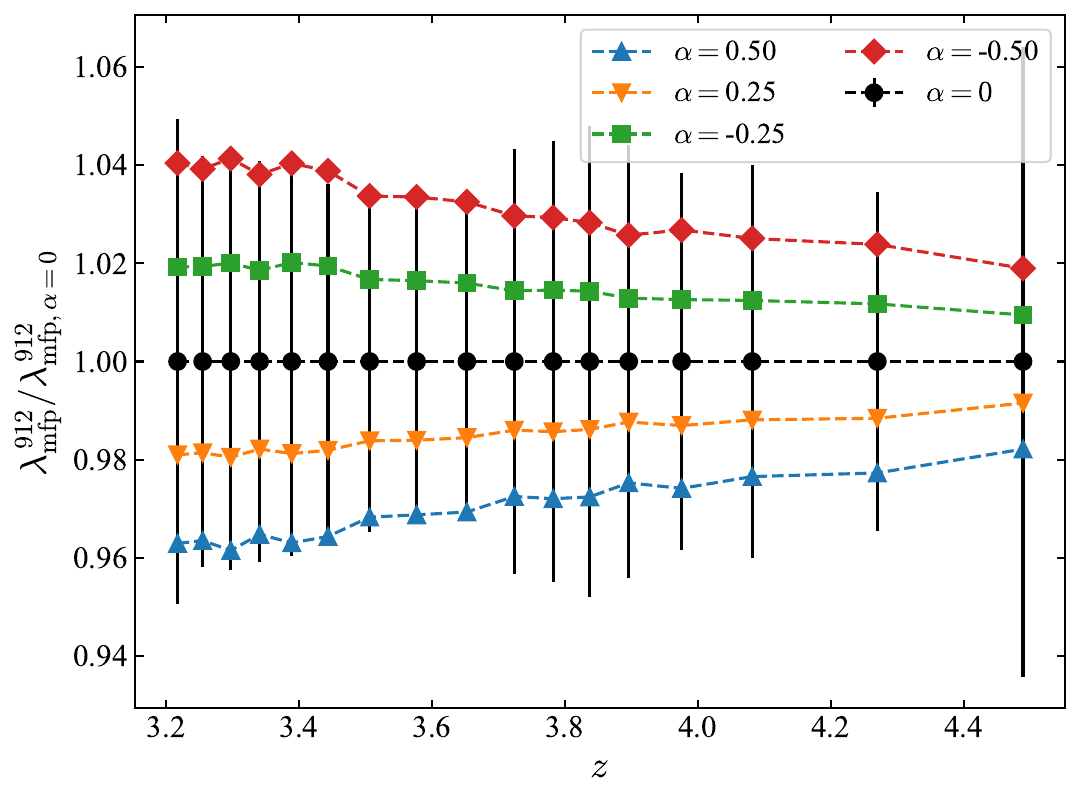}
    \end{minipage}\hfill
    \begin{minipage}{\columnwidth}
        \centering
        \includegraphics[width=\columnwidth]{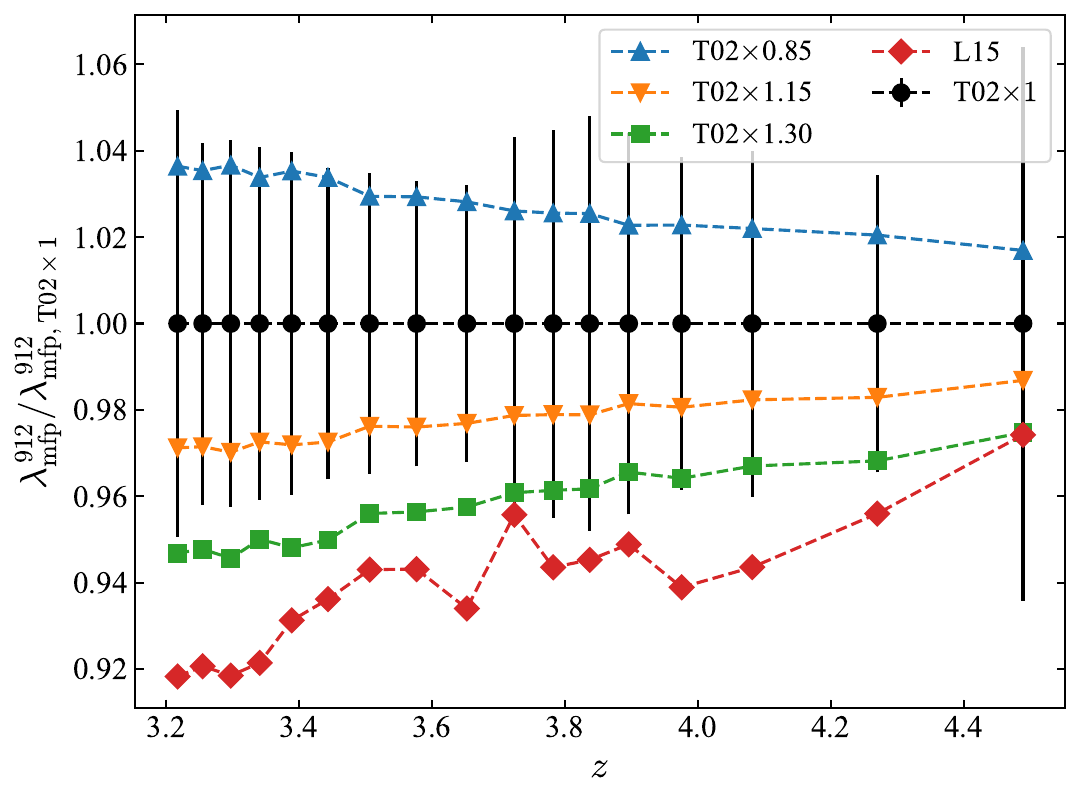}      
    \end{minipage}
    \caption{Measurements of different quasar SEDs compared with our default model. Left panel: Measurements when we adjust the tilt index $\alpha$ in \autoref{eq5} to 0.5 (blur triangle-ups), 0.25 (orange triangle-downs), -0.25 (green squares) and -0.5 (red hexagons), which show that except for an extreme tilt ($|\alpha|\gtrsim0.5$), our default measurements would not bias for more than 5\% due to an incorrect spectral index. Right panel: Measurements with different overall normalization. The blue triangle-ups, orange triangle-downs and green squares represent the measurements when we shift the \citetalias{2002ApJ...565..773T} SED to its 0.85, 1.15 and 1.30 times. For comparison we also draw the measurements with \citetalias{2015MNRAS.449.4204L}, represented by red hexagons. The difference between \citetalias{2015MNRAS.449.4204L} and \citetalias{2002ApJ...565..773T} measurements can then be explained for $\sim 70\%$ by the overall height of SED, and the remaining $\sim 30\%$ may come from the shape fluctuations.}
    \label{fig7}
\end{figure*}

One of the most prominent sources of systematics come from the choice of quasar SED, which is highly influenced by sample selection and the method of IGM absorption correction. Parametrizing SED as $f_\nu\propto\nu^{\alpha_\nu}$, various studies show a similar spectral index $\alpha_\nu$ in the Lyman continuum region. The quasar spectrum composite from \citetalias{2002ApJ...565..773T} with a sample of 184 quasars and a median redshift $z\simeq1.2$ has a spectral index $\alpha_\nu=-1.76\pm0.12$ at wavelength $500\sim1200$\AA, and \citetalias{2015MNRAS.449.4204L} measured a similar result $\alpha_\nu=-1.70\pm0.61$ at $z\simeq2.4$. Other measurements, however, give different indices. \citet{2014ApJ...794...75S} exploits the spectra of 159 active galactic nuclei (AGNs) at a median redshift $z\simeq0.37$ observed with \textit{HST}/COS and give $\alpha_\nu=-1.41\pm0.15$. \citet{2023NatAs...7.1506C} reports a more extreme spectral index $\alpha_\nu=-2.4\pm0.1$ after correcting the quasar selection bias. Thus it is necessary to assess the systematic bias induced by the uncertainty of spectral index. By altering the tilt factor $\alpha$ in \autoref{eq5}, we can change this spectral index without changing the overall level of SED. We show our measurements with different $\alpha$ in the left panel of \autoref{fig7}, which indicates that unless the true underlying SED exhibits an extreme spectral index and differs 
by more than 0.5, our measurements would not be biased by more than 5\%, which is of the same scale of the statistical uncertainty, as shown in \autoref{fig7}. It is then clear that if we want to explain the discrepancy with SED slope, we would have to choose a slope that will contradict with most of previous measurements.

Another source of systematics is the overall level of SED at the Lyman continuum. Although \citetalias{2002ApJ...565..773T} and \citetalias{2015MNRAS.449.4204L} have nearly the same spectral index, their overall level differs 
by $\sim30\%$ in the Ly$\alpha$ forest when
normalized at $1450$\AA, as shown in \autoref{fig2}. Other SED templates have the same problem, summarized in Figure 8 of \citetalias{2015MNRAS.449.4204L} and Figure 3 of \citet{2023NatAs...7.1506C}. This problem mainly comes from the IGM absorption correction, which 
depends highly on the form of \specnotation{H}{I} absorber distribution function $f(N_{\specnotation{H}{I}},z)$. \citetalias{2015MNRAS.449.4204L} choose the \citet{2014MNRAS.438..476P} spline model, and \citetalias{2002ApJ...565..773T} choose an empirical broken power law with parameters from various studies (see their Section 2.2). The effect of SED overall level is reflected in the estimation of $\tau^\mathrm{LyS}_0$ in \autoref{eq6}, where a higher level will result in a larger $\tau^\mathrm{LyS}_0$ and therefore a faster Lyman series opacity evolution. The right panel of \autoref{fig7} shows our measurements with \citetalias{2002ApJ...565..773T} SED multiplied by different normalization factors. The result is consistent with our thought: A higher SED will bring a faster declining of Lyman series opacity from restframe Lyman limit to shorter wavelengths, requiring the Lyman limit opacity to be higher to compensate for the total absorption. The differences between measurements using \citetalias{2002ApJ...565..773T} and \citetalias{2015MNRAS.449.4204L} can then be mostly explained by the higher overall level of \citetalias{2015MNRAS.449.4204L} SED, and the remaining difference may come from the different SED fluctuations, e.g. the bump at restframe $\sim 800$\AA\;in \citetalias{2015MNRAS.449.4204L} SED shown in \autoref{fig2}. Combining the two uncertainties in quasar SED, we believe that the systematic bias induced by SED would not exceed $\sim 10\%$, which means the SED choice cannot entirely
explain the $\sim25\%$ discrepancy with \citetalias{2014MNRAS.445.1745W}.

Neglecting the Lyman limit opacity evolution when fitting, i.e. enforcing $\gamma_\kappa=0$ in \autoref{eq12} may also introduce bias to our model. Most of the previous work at similar redshifts did not consider this evolution (\citetalias{2009ApJ...705L.113P}, \citealt{2013ApJ...775...78F}, \citetalias{2014MNRAS.445.1745W}, \citealt{2018ApJ...860...41L}). \citet{2013ApJ...765..137O} fitted this parameter and found a mean value $\gamma_\kappa\simeq0.4$ and a median value $\gamma_\kappa\simeq0$ at $z\simeq2$, which means $\gamma_\kappa=0$ would be a suitable approximation. To better explore the influence of freeing $\gamma_\kappa$ on the mean free path measurements, we use the untilted \citetalias{2002ApJ...565..773T} SED to fit our stacked spectra with a $\gamma_\kappa>0$ uniform prior. These measurements are shown in \autoref{tab3}, together with an example combined MCMC posterior shown in {the lower left panel of \autoref{figMC}}, where the results are mostly consistent with results in \autoref{tab1}, except that the result of the lowest redshift in \autoref{tab3} is not physical, which we believe comes from a wavelength range that is too short to constrain $\gamma_\kappa$. Additionally, most $\gamma_\kappa$ fittings also deviate from 0, which is not surprising as the mean free path also evolves quickly in these redshift. In conclusion, neglecting $\gamma_\kappa$ would not bias our measurements, so we continue to keep $\gamma_\kappa=0$ hereafter. 

\begin{table}
  \centering
  \caption{$\gamma_\kappa$ and $\lambda^{912}_\mathrm{mfp}$ measurements when freeing $\gamma_\kappa$. The unit of $\lambda^{912}_\mathrm{mfp}$ is $h_{70}^{-1}\,\mathrm{Mpc}$. Since we applied $\gamma_\kappa>0$ prior, the $\gamma_\kappa$ estimates with their posterior peaking at $\gamma_\kappa=0$ are underlined, and their uncertainties are the $50\%$ quantile. Other posteriors peak at nearly $50\%$, so the upper and lower uncertainties are chosen to be the $84\%$ and $16\%$ quantiles.}
    \begin{tabular}{ccc}
    \toprule
    $z_\mathrm{median}$  & $\gamma_\kappa$ & $\lambda^{912}_\mathrm{mfp}$ \\
    \midrule
    3.217 & $11.56^{+3.79}_{-3.67}$ & $439.4\pm5.88$ \\
    3.255 & \underline{$0.00+0.68$} & $65.4\pm3.10$  \\
    3.297 & \underline{$0.00+2.28$} & $68.1\pm3.45$ \\
    3.341 & $3.81^{+1.92}_{-1.77}$ & $61.0\pm2.93$ \\
    3.389 & $3.07^{+1.75}_{-1.65}$ & $63.9\pm3.01$ \\
    3.443 & $3.82^{+1.46}_{-1.45}$ & $60.3\pm2.71$ \\
    3.506 & $2.82^{+1.16}_{-1.23}$ & $50.9\pm2.14$ \\
    3.577 & $2.29^{+1.08}_{-1.14}$ & $49.2\pm1.91$ \\
    3.653 & $2.93^{+1.11}_{-1.07}$ & $45.3\pm1.83$ \\
    3.724 & $3.72^{+1.38}_{-1.45}$ & $41.2\pm2.22$ \\
    3.782 & $2.53^{+1.45}_{-1.42}$ & $39.5\pm2.17$ \\
    3.837 & $4.02^{+1.30}_{-1.29}$ & $38.1\pm2.20$ \\
    3.896 & $2.22^{+1.36}_{-1.29}$ & $33.9\pm1.87$ \\
    3.976 & \underline{$0.00+0.99$} & $35.1\pm1.64$ \\
    4.082 & $1.69^{+1.26}_{-1.12}$ & $30.5\pm1.53$ \\
    4.270 & \underline{$0.00+0.50$} & $29.1\pm1.22$ \\
    4.489 & \underline{$0.00+3.25$} & $22.1\pm2.06$ \\
    \bottomrule
    \end{tabular}%
  \label{tab3}%
\end{table}%

\begin{figure*}[htbp]
    \centering
    \includegraphics[width=2\columnwidth]{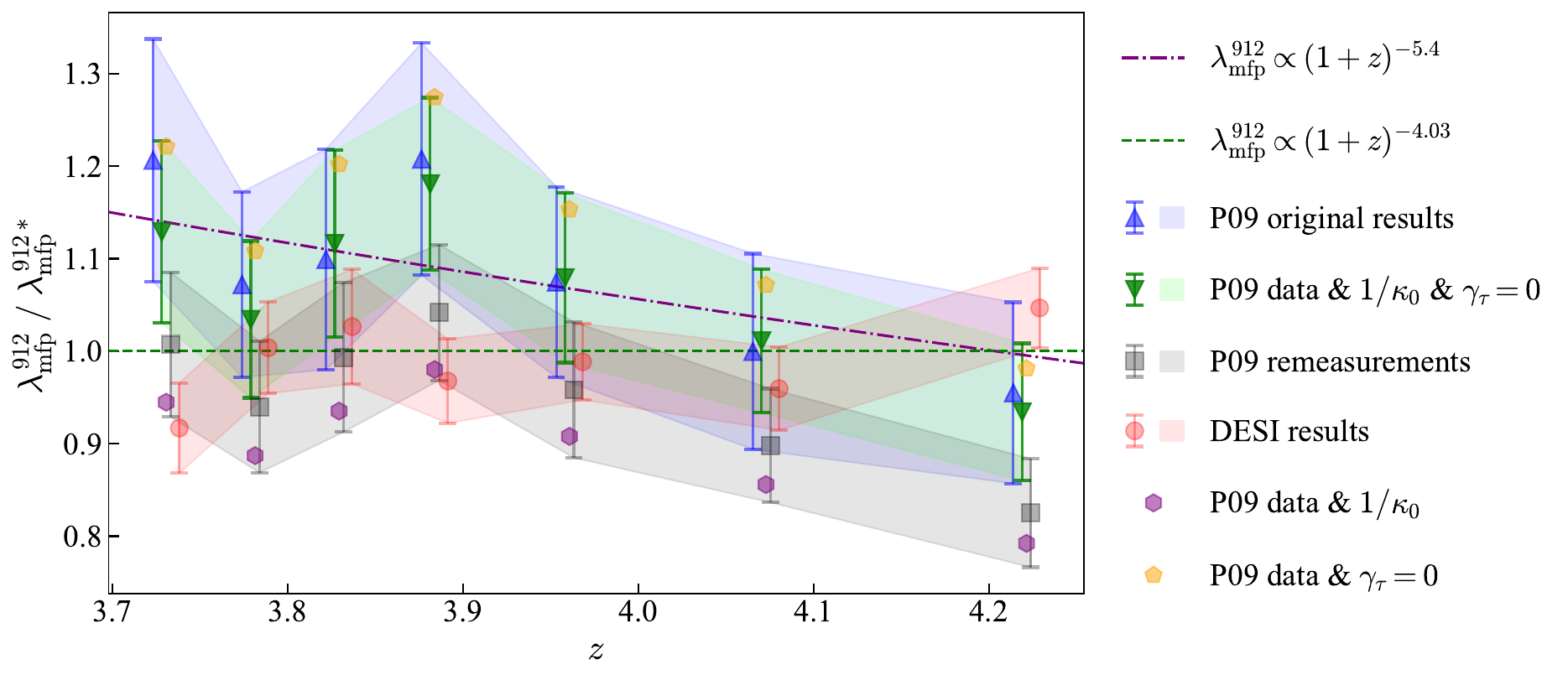}
    \caption{Comparison between different $\lambda_\mathrm{mfp}^{912}$ measurements. All the points and lines are normalized with our \citetalias{2002ApJ...565..773T} power law fitting in \autoref{tab1} (denoted as $\lambda_\mathrm{mfp}^{912*}$ in $y$-axis label) {and are slightly shifted along the $x$-axis for a better display}. The purple dash-dot line and green dash line are the \citetalias{2014MNRAS.445.1745W} power law model and our $\lambda_\mathrm{mfp}^{912*}$ model. The blue up-triangles are the original \citetalias{2009ApJ...705L.113P} results. With the exactly same data, we use our method (grey squares), our method with $\lambda_\mathrm{mfp}^{912}$ definition to be $1/\kappa_0$ (purple pentagons), our method with $\gamma_\tau=0$ (orange hexagons) and our method with both $1/\kappa_0$ definition and $\gamma_\tau=0$ (green down-triangles) to measure $\lambda^{912}_\mathrm{mfp}$, respectively. Measurements with DESI stacks at the same redshift range are shown in red circles for comparison. To better illustrate the uncertainty estimates, the blue, green, grey and red dots are shaded with the same color within $1\sigma$ range, while the purple and orange dots are not. Comparing blue and green triangles, we believe that our method stays consistent with previous works in estimating both values and uncertainties. Correcting the $\gamma_\tau$ choice brings the \citetalias{2009ApJ...705L.113P} measurements consistent with ours (grey squares), while the $z=4.07$ and $z=4.22$ results deviates from our $\lambda_\mathrm{mfp}^{912*}$ model for more than $1\sigma$. The different $\lambda_\mathrm{mfp}^{912}$ definition would generate $\sim7\%$ bias, and neglecting $\gamma_\tau$ would bias the result for $\sim20\%$. }
    \label{fig9}
\end{figure*}

After the above considerations, we believe that the reason why our measurements are systematically lower than the power law in \citetalias{2014MNRAS.445.1745W}, or \citetalias{2009ApJ...705L.113P} more specifically, is the neglection of Lyman series opacity \textit{evolution} in \citetalias{2009ApJ...705L.113P}. In other words, \citetalias{2009ApJ...705L.113P} simply replace the Lyman series opacity at the blueward of 912\AA\;with $\tau_0^\mathrm{LyS}$ (\autoref{eq6}) that is estimated at exactly 912\AA. By this means they overestimated the Lyman series opacity, leading to an overestimation of the mean free path. To test our idea, we use exactly the same data in \citetalias{2009ApJ...705L.113P} and add the $\gamma_\tau=3$ evolution to remeasure the mean free path, shown in \autoref{fig9} (purple pentagons), together with the original value in \citetalias{2009ApJ...705L.113P} (blue up-triangles). Comparing the two sets of data, the neglection of the Lyman limit opacity evolution could produce a discrepancy more than $20\%$, implying this is a major source of the difference. However, correcting this neglection would over-correct the mean free path results and make them lower than the $\lambda_\mathrm{mfp}^{912}\propto(1+z)^{-4.03}$ model (\autoref{fig9}, purple pentagons v.s. green dash line), which indicates the existence of other bias.

We believe the last source of systematic bias is the definition of the mean free path. Different from ours, \citetalias{2009ApJ...705L.113P} defines $\lambda_\mathrm{mfp}^{912}=1/\kappa_0$, where $\kappa_0$ is the opacity at the redshift of emission quasars (\autoref{eq9}).  Theoretically, the $1/\kappa_0$ definition estimates the opacity at the light emission source, while the $\tau^{\mathrm{LyC}}=1$ definition involves the opacity evolution along the travel of the photons, which would take the evolved lower opacity into consideration and generate a higher mean free path. Results shown in \autoref{fig9} support this claim: measurements with $1/\kappa_0$ definition are indeed lower by $5-10\%$ than that with the $\tau^\mathrm{LyC}=1$ definition (orange hexagons v.s. green down-triangles, grey squares v.s. purple pentagons). In \citet{2014MNRAS.442.1805I}, the authors also explored the difference between different definitions and came to a similar conclusion. Interestingly, the definition choice in \citetalias{2009ApJ...705L.113P} partially offsets the impact of setting $\gamma_\tau=0$, and together they produce the $\sim20\%$ discrepancy in \citetalias{2009ApJ...705L.113P} redshift range. Taking the two factors into consideration, our fitting pipeline reproduces the results in \citetalias{2009ApJ...705L.113P} (\autoref{fig9}, green down-triangles), which further confirms the validity of our method.

To sum up, we assess several possible sources of systematic bias in the mean free path measurements. The uncertainty of quasar SED, including its spectral index and the overall level at the Lyman continuum, would bias the mean free path results for $\sim10\%$. Given that the precise quasar SED is still under debate, a less biased SED template is needed for future mean free path measurements. The neglecting of $\gamma_\kappa$ do not introduce bias and is appropriate for most redshift ranges, but the different treatment of $\gamma_\tau$ is the key factor that causes the discrepancy in \autoref{fig5}.
Different mean free path definitions would also introduce secondary bias into our model. {Note that although we have identified the main sources of discrepancy, \autoref{fig9} still shows a slight discrepancy between \citetalias{2009ApJ...705L.113P} original results and our re-measurements at the low redshift end. We believe it is from other smaller differences in our fitting method, e.g. the treatment of flux uncertainties, or as
yet unidentified systematic errors.}

\subsection{Revised Redshift Evolution of the Mean Free Path}\label{sec43}

Combining the direct measurements from \citet{2013ApJ...765..137O}, \citet{2013ApJ...775...78F}, \citetalias{2014MNRAS.445.1745W}, \citet{2018ApJ...860...41L}, \citet{2021MNRAS.508.1853B}, \citet{2023ApJ...955..115Z}, remeasured \citetalias{2009ApJ...705L.113P} and our result using \citetalias{2002ApJ...565..773T} SED, we suggest a broken power law parameterization of the redshift evolution of mean free path, described by:
\begin{equation}
    \lambda_\mathrm{mfp}^{912}=\begin{cases}
        \displaystyle A\left(\frac{1+z}{1+z_0}\right)^{\eta_1},\quad\mathrm{when}\;z\leq z_0\\[10pt]
        \displaystyle A\left(\frac{1+z}{1+z_0}\right)^{\eta_2},\quad\mathrm{when}\;z>z_0\\
    \end{cases}
    \label{eq15}
\end{equation}
where $A, z_0, \eta_1, \eta_2$ are all free parameters. {Note that there has not been a satisfying and theoretically-inspired functional form to describe the \lmfp \,\,evolution at redshift $z\gtrsim5$, and our choice here is empirical. Our main purpose is to give a constraint for the transition redshift $z_0$, instead of giving a precise description of \lmfp \,\,evolution at $z\gtrsim5$.}

\begin{figure}
    \centering
    \includegraphics[width=\columnwidth]{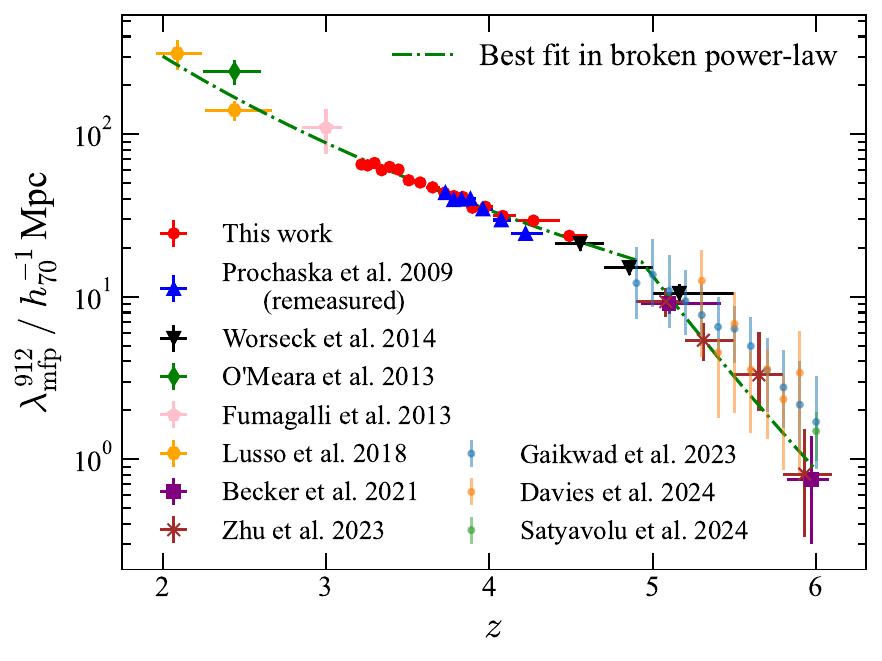}
    \caption{The {best fitted} broken power law of the current mean free path measurements at $2\lesssim z\lesssim6$, using the functional form in \autoref{eq15}. The data points in \citetalias{2009ApJ...705L.113P} (blue up-triangles) are remeasured using our pipeline, as what we do in \autoref{fig9}.}
    \label{fig10}
\end{figure}

With the assumption of Gaussian uncertainty, the best parameter estimates given by MCMC posterior are $A=16.51^{+1.85}_{-0.91}\,h_{70}^{-1}\,\mathrm{Mpc},\;z_0=4.93^{+0.06}_{-0.12},\;\eta_1=-4.27^{+0.14}_{-0.12},\;\eta_2=-17.73^{+3.13}_{-3.52}$. \autoref{fig10} shows the well-behaved MCMC posterior and the best fitted line that is roughly consistent with all the existing direct measurements of mean free path. Measurements in \citet{2023MNRAS.525.4093G}, \citet{2024ApJ...965..134D} and \citet{2024MNRAS.533..676S} in \autoref{fig10} are not included in the fitting, which we will discuss in Appendix~\ref{app3}. As the parameter posterior is not Gaussian, we choose the maximum posterior estimates, and the errorbars indicate the \textit{scaled} $1\sigma$ range (e.g. if the best estimate corresponds to $40\%$ quantile, then the lower errorbar represent the $(40\times16/50)\%$ quantile, and the upper errorbar is the $(1-60\times16/50)\%$ quantile). We stress that we do not correct for the different SEDs of previous works in this fitting: \citetalias{2009ApJ...705L.113P}, \citet{2013ApJ...765..137O}, \citet{2013ApJ...775...78F}, \citet{2014MNRAS.445.1745W} and our results use the \citetalias{2002ApJ...565..773T} SED, while \citet{2018ApJ...860...41L}, \citet{2021MNRAS.508.1853B} and \citet{2023ApJ...955..115Z} use the variations of \citetalias{2015MNRAS.449.4204L} SED. But the fitting would not bias significantly, especially at higher redshift, as we can see that the three independent measurements from \citetalias{2014MNRAS.445.1745W}, \citet{2021MNRAS.508.1853B} and \citet{2023ApJ...955..115Z} at $z\simeq5$ almost coincide with each other. The posterior shows a distinct double-peak feature, but estimates from the two peaks are almost identical and only differ at $z\sim z_0$, so we keep the estimate from the highest peak. We calculate the Bayesian information criterion (BIC) of our broken power law model and single power law model, defined as:
\begin{equation}
    \mathrm{BIC}=\chi^2+p\ln{n}
\end{equation}
where $n$ is the number of data points, 37 in this case, and $p$ is the number of parameters, 4 for broken power law and 2 for single power law. Note that the $\chi^2$ here is computed among the $\lmfp$ measurements and is \textit{not} the one defined in \autoref{eq11}. Results show that the broken power law model achieves $\mathrm{BIC}=51.0$, much lower than the single power law model $\mathrm{BIC}=330.7$. Our fitting shows that after correcting all the systematics, a slower evolution of mean free path after $z\simeq5$ and a break redshift $z_0=4.93$ are strongly favored by the existing data. 

\section{Discussion} \label{sec5}

\subsection{Impact on the Estimation of $f(N_{\specnotation{H}{I}},z)$}

\begin{figure}
    \centering
    \includegraphics[width=\columnwidth]{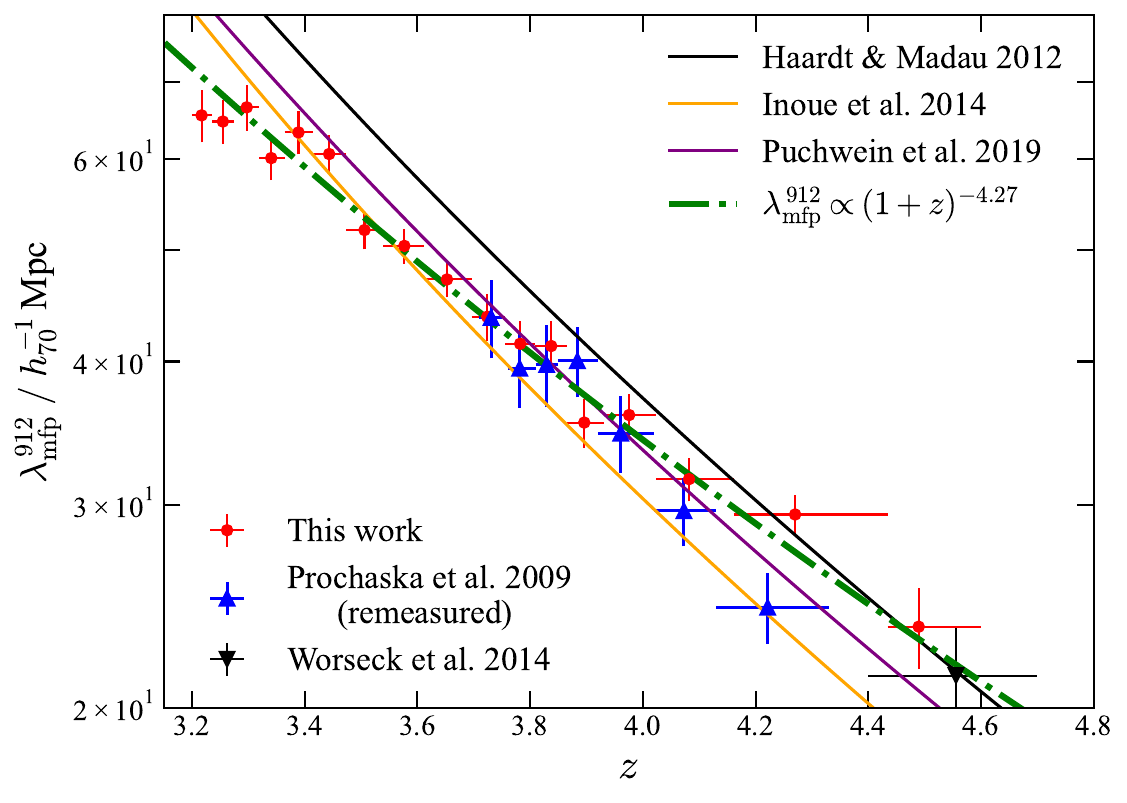}
    \caption{Comparison between our measurements and the previous $f(N_{\specnotation{H}{I}},z)$ models. The red circles, blue up-triangles and black down-triangles are the observational measurements from our work, remeasured \citetalias{2009ApJ...705L.113P} and \citetalias{2014MNRAS.445.1745W}, respectively. The black, purple and yellow lines are the mean free path calculations with the \autoref{eq10} definition, using models from \citet{2012ApJ...746..125H}, \citet{2014MNRAS.442.1805I} and \citet{2019MNRAS.485...47P}, respectively. The green dash-dot line represents our best broken power law fitting. All the three models in literature over-predict the evolution speed of the mean free path.}
    \label{fig11}
\end{figure}

Many empirical estimates of $f(N_{\specnotation{H}{I}},z)$ have been proposed to study the IGM opacity and provide a theoretical ingredient to calculate the EUVB. We compare the measurements in this work with the most widely used models in the literature: \citet{2012ApJ...746..125H}, \citet{2014MNRAS.442.1805I} and \citet{2019MNRAS.485...47P} in \autoref{fig11}, with the $z_{912}$ for calculating $\lmfp$ using \autoref{eq10} defined as:
\begin{align}
\begin{split}
        \tau^\mathrm{LyC}(z_{912}, z_\mathrm{qso})=&\int_0^{\infty}\,\mathrm{d}N_{\specnotation{H}{I}}\int_{z_\mathrm{912}}^{z_\mathrm{qso}}f(N_{\specnotation{H}{I}},z)\\
    &\times\{1-\exp\left[-N_{\specnotation{H}{I}}\sigma_\mathrm{ph}(z)\right]\}\,\mathrm{d}z=1\label{eq14}\\
\end{split}
\end{align}
where $\sigma_\mathrm{ph}(z)$ is the hydrogen photoionization cross section when the photon reaches redshift $z$. We do not include the model in \citet{2020MNRAS.493.1614F}, as it only differs from \citet{2019MNRAS.485...47P} at low redshift. \citet{2019MNRAS.485...47P} (purple line) best matches our results among the three models, while \citet{2012ApJ...746..125H} overpredict the mean free path at $z\gtrsim 3.2$. { \citet{2014MNRAS.442.1805I} underpredict the mean free path at $z\gtrsim4$ by $\sim15\%$, with a $p$-value from $\chi^2$-goodness-of-fit test to be $\sim1.5\times10^{-8}$.} Nevertheless, the three models all overpredict the evolution speed implied by our measurements, illustrated with the green dash-dot line. However, the functional form in \citet{2010ApJ...721.1448S} coincides with our results. All these models are derived from the Ly$\alpha$ absorber statistics, which are easily biased by e.g. line blending and the clustering of absorption systems \citep{2010ApJ...718..392P, 2014MNRAS.438..476P}, while our method does not depend on the specific distribution of Ly$\alpha$ absorbers. Therefore, we encourage a further exploration of systematics and a modification of the form of $f(N_{\specnotation{H}{I}},z)$ to better match the observations.

\subsection{Implication for Reionization}

The history of reionization has been studied by a large body of literature with various methods. Measurements of the optical depth for the Cosmic Microwave Background (CMB) photons set a midpoint for the reionization $z_\mathrm{re}=7.82\pm0.71$ \citep{2021A&A...652C...4P}. Although it was once believed that reionization ends at $z\simeq6$ (e.g. \citealt{2015MNRAS.447..499M}), recent observations of high redshift quasar spectra provide evidence that reionization may extend significantly below $z\simeq6$. The long dark gap statistics in Ly$\alpha$ and Ly$\beta$ forests strongly rule out a scenario with homogeneous UV background and a reionization fully completed at $z>6$ \citep{2021ApJ...923..223Z, 2022ApJ...932...76Z}. Measurements of the Ly$\alpha$ effective optical depth and its fluctuations further confirm the judgement and imply an ending that could be as late as $z\simeq5.3$ (e.g. \citealt{2019MNRAS.485L..24K, 2020ApJ...904...26Y, 2021MNRAS.501.5782C, 2022MNRAS.514...55B}). The methods of dark pixels (e.g. \citealt{2023ApJ...942...59J}) and damping wings (e.g. \citealt{2024ApJ...969..162D, 2024MNRAS.533L..49Z, 2024A&A...688L..26S, 2024MNRAS.530.3208G}) are used to constrain the neutral fraction of IGM at $5\lesssim z\lesssim6$ with in Ly$\alpha$ forest, which is a direct probe of the reionization history.

The measurements of mean free path could also be an independent probe for reionization. The mean free path is first controlled by the typical size of \specnotation{H}{II} bubbles and evolves rapidly at the beginning of reionization. After the overlapping of these bubbles, the mean free path would then be controlled by the optically-thick LLS \citep{2017ApJ...851...50M, 2022LRCA....8....3G}. The redshift evolution of the mean free path can therefore be separated into two stages, which is indeed shown by our broken power law fitting in \autoref{eq15} and reproduced in numerical simulations (e.g. \citealt{2021ApJ...917L..37C, 2022MNRAS.516.3389L}). However, the transition redshift $z_0=4.93$ that is supposed to be the time of overlapping is later than most previous estimates where the neutral fraction of IGM is constrained to be $\sim10^{-4}$ at $z\sim5.5$ \citep{2023ARA&A..61..373F}. This mild tension may come from the inappropriate choice of our functional form at $z\gtrsim z_0$, but it is difficult for a more precise model to generate a more consistent transition redshift given the lack of mean free path measurements at $z\gtrsim5$ and the large uncertainty of the existing results. Nevertheless, the measurements of Ly$\alpha$ effective optical depth $\tau_\mathrm{eff}$ may provide a supplement support for our claim. The power law evolution of $\tau_\mathrm{eff}$ at $2\lesssim z\lesssim5$ has an index $\beta\simeq3$ (e.g. \citealt{2008ApJ...681..831F, 2013MNRAS.430.2067B, 2020ApJ...892...70K, 2021MNRAS.506.4389G, 2024MNRAS.532.2082D, 2024arXiv240506743T}), while at $z\gtrsim5$ 
the evolution becomes much faster, with $\beta\simeq10$ (e.g. \citealt{2020ApJ...904...26Y, 2022MNRAS.514...55B}). This roughly coincides with our $\lmfp$ evolution in both the break redshift and the power law index at either sides. In conclusion, the break of $\lmfp$ evolution at $z=z_0=4.93$ implies that reionization may end at this time, but further evidence and interpretation are still needed to better align with other independent measurements.

\section{Conclusion} \label{sec6}

We have presented new measurements of the mean free path of neutral hydrogen ionizing photons at $3.2\leq z\leq4.6$. We construct the largest quasar sample containing 12,595 quasars from DESI Y1 observations. By fitting the Lyman continuum of the composite quasar spectra at 17 redshift bins, we obtain the most precise measurements to date, summarized in Table~\ref{tab1}. Our results are systematically lower at $z\gtrsim 3$ than the previous power law interpolation in \citetalias{2014MNRAS.445.1745W} and show a much shallower evolution speed. We investigate a number of possible biases, and conclude that the main reason is the neglection of the evolution of Lyman series opacity in \citetalias{2009ApJ...705L.113P}. Combining with all the previous estimates, we believe that the redshift evolution can be best described by a broken power law, with the break redshift $z_0=4.93^{+0.06}_{-0.12}$. This evolution is slower than the predictions from most of the current $f(N_{\specnotation{H}{I}},z)$ models, and although later than most other measurements, the power law break at $z\approx5$ may be interpreted as the end of \specnotation{H}{I} reionization. As even larger quasar samples from DESI is coming in the next few years, more precise measurements of $\lmfp$ as well as other quantities such as $\tau_\mathrm{eff}$ can be conducted in the future, which we believe will provide tighter constraints on the thermal evolution of IGM at post-reionization era.

\begin{acknowledgments}

We thank Piero Madau and Frederick B. Davies for their constructive advice. We thank Ting-Wen Lan and Vid Irsic for their helpful comments during the DESI internal review. SZ acknowledges support from the National Science Foundation of China (no. 12303011). 

This research is based upon work supported by the U.S. Department of Energy (DOE), Office of Science, Office of High-Energy Physics, under Contract No. DE–AC02–05CH11231, and by the National Energy Research Scientific Computing Center, a DOE Office of Science User Facility under the same contract. Additional support for DESI was provided by the U.S. National Science Foundation (NSF), Division of Astronomical Sciences under Contract No. AST-0950945 to the NSF’s National Optical-Infrared Astronomy Research Laboratory; the Science and Technology Facilities Council of the United Kingdom; the Gordon and Betty Moore Foundation; the Heising-Simons Foundation; the French Alternative Energies and Atomic Energy Commission (CEA); the National Council of Humanities, Science and Technology of Mexico (CONAHCYT); the Ministry of Science, Innovation and Universities of Spain (MICIU/AEI/10.13039/501100011033), and by the DESI Member Institutions: \url{https://www.desi.lbl.gov/collaborating-institutions}. Any opinions, findings, and conclusions or recommendations expressed in this material are those of the author(s) and do not necessarily reflect the views of the U. S. National Science Foundation, the U. S. Department of Energy, or any of the listed funding agencies.

The authors are honored to be permitted to conduct scientific research on Iolkam Du’ag (Kitt Peak), a mountain with particular significance to the Tohono O’odham Nation.

Data for tables and figures and the code we used are both avilable in Zenodo at https://doi.org/10.5281/zenodo.14167714.

\end{acknowledgments}

% \vspace{5mm}

% \software{astropy \citep{2013A&A...558A..33A, 2022ApJ...935..167A}, numpy \citep{2011CSE....13b..22V}, matplotlib \citep{2007CSE.....9...90H}, scipy \citep{2020NatMe..17..261V}, emcee \citep{2013PASP..125..306F}, corner \citep{2016JOSS....1...24F}}

\appendix

\section{Examples of Spectra} \label{app0}

\begin{figure*}
    \centering
    \begin{minipage}{0.49\columnwidth}
        \centering
        \includegraphics[width=\columnwidth]{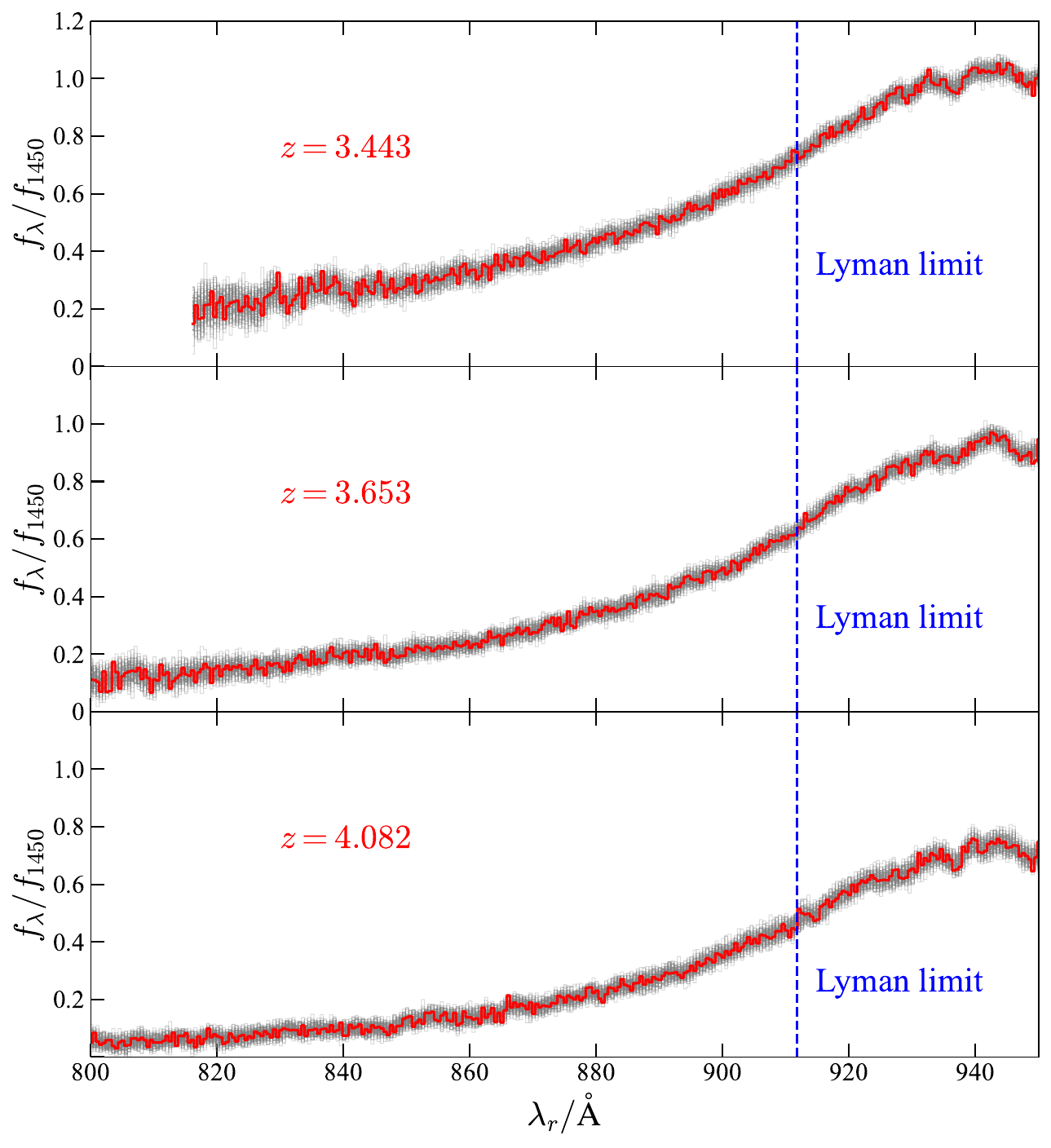}
    \end{minipage}\hfill
    \begin{minipage}{0.49\columnwidth}
        \centering
        \includegraphics[width=\columnwidth]{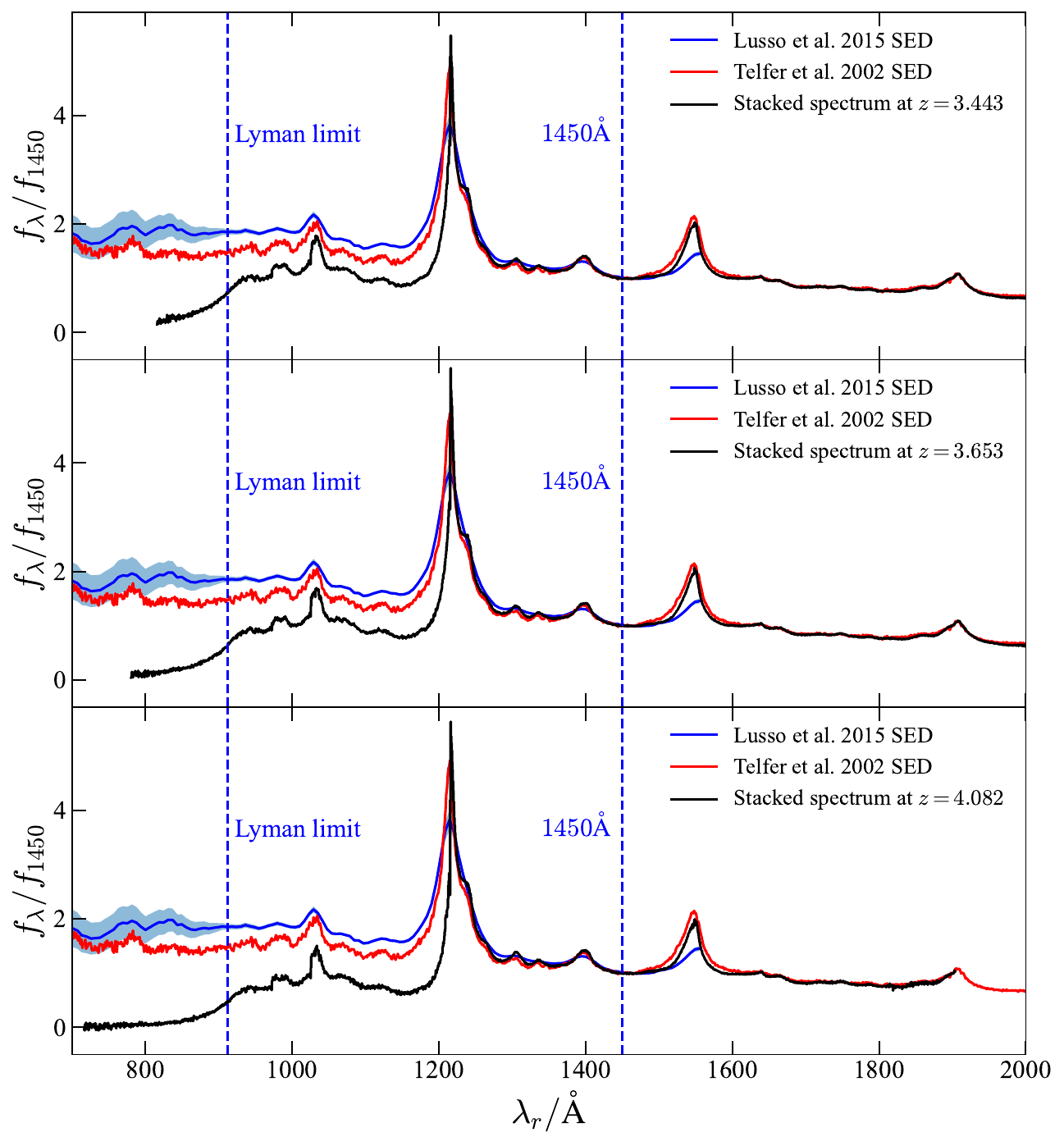}
    \end{minipage}
    \caption{Left panel: Examples of the composite spectra from bootstrap technique (grey), together with the composite spectra from the normal stacking procedure (red). Right panel: Comparisons between the SED templates and our composite spectra at $700\mathrm{\AA}\leq\lambda_r\leq2000\mathrm{\AA}$.}
    \label{figZ}
\end{figure*}

To better illustrate the uncertainty level of our composite spectra, we show examples of the bootstrap spectra in \autoref{figZ}. The redshift is chosen to be the same as \autoref{fig2}. We also show the spectra redward of Lyman$\alpha$ at the right panels of \autoref{figZ} together with SED templates.

\section{Lyman Series Opacity Evolution}\label{app2}

\begin{figure}
    \centering
    \includegraphics[width=0.5\columnwidth]{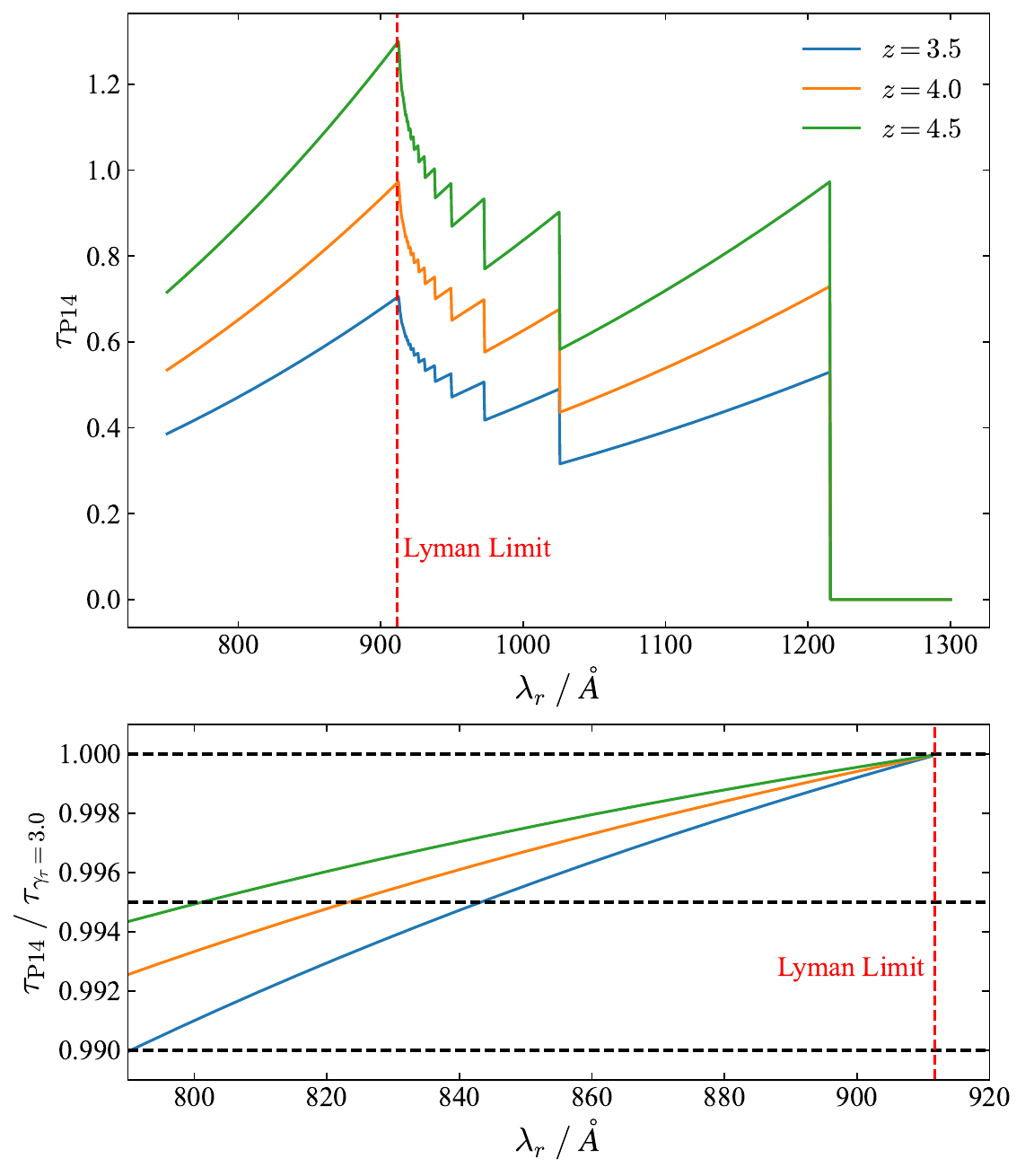}
    \caption{The Lyman series optical depth calculated using the $f(N_{\specnotation{H}{I}}, z)$ model in \citet{2014MNRAS.438..476P} (P14), at the restframe of three $z=3.5$ (blue), $4.0$ (orange) and $4.5$ (green) quasars. The upper panel shows the optical depth accounting for all Lyman series lines. The lower panel shows the ratio of the optical depth calculated from P14 model and our \autoref{eq6}, with the black horizontal dash lines representing 1.000, 0.995 and 0.990 and the red vertical dash line representing the Lyman limit wavelength (912\AA). The Lyman series opacity at the blueward of the Lyman limit wavelength clearly evolves, and our approximation of Lyman series optical depth in \autoref{eq6} is shown to produce only $<1$\% systematic error, and for most cases $<0.5$\%.}
    \label{fig3}
\end{figure}

In \autoref{eq6}, we approximate the Lyman series opacity using a power law and the index $\gamma_\tau$ is chosen to be $3$. \autoref{fig3} shows the comparison between our $\gamma_\tau=3$ approximation and the Lyman series opacity calculated with \citet{2014MNRAS.438..476P} model, which shows only $\lesssim1\%$ difference, indicating that $\gamma_\tau=3$ is a suitable approximation. 

\section{{Corner plots of MCMC posteriors}} \label{appMC}

\begin{figure*}[htbp]
    \centering
    % \begin{minipage}{0.48\columnwidth}
    %     \centering
    %     \includegraphics[width=\columnwidth]{corner.pdf}      
    % \end{minipage}\hfill
    % \begin{minipage}{0.48\columnwidth}
    %     \centering
    %     \includegraphics[width=\columnwidth]{mfp_hist.pdf}
    % \end{minipage}

    \begin{tabular}{cc}
  \includegraphics[width=0.48\columnwidth]{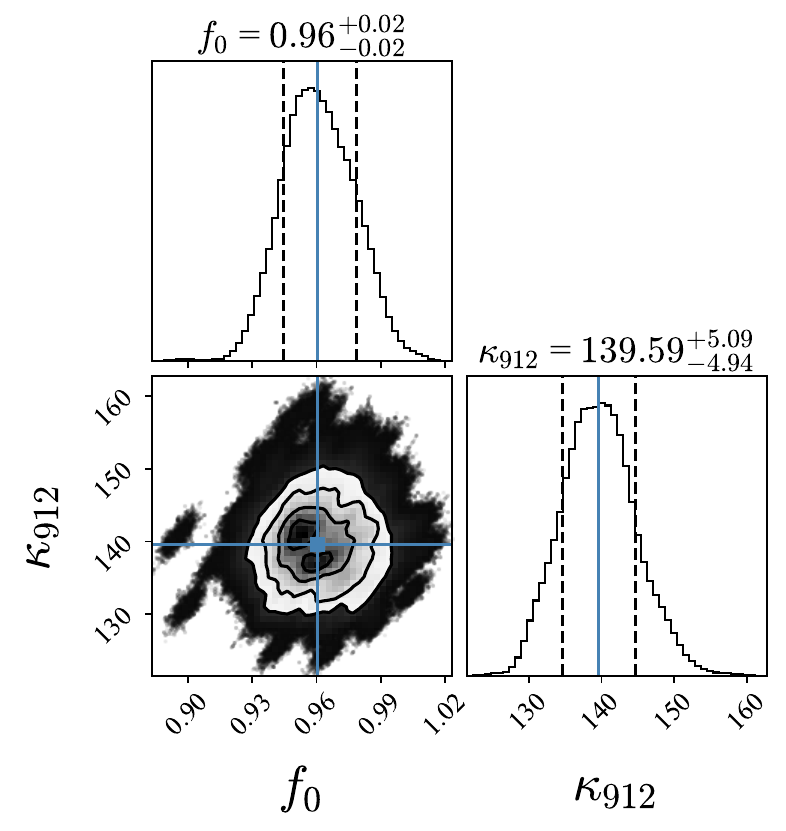} &   \includegraphics[width=0.48\columnwidth]{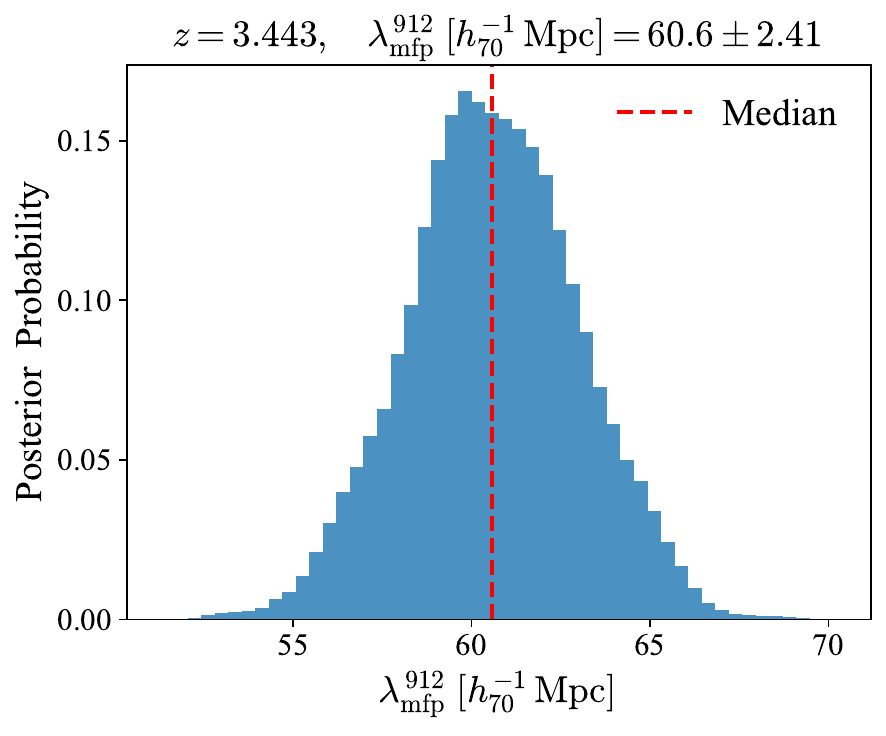} \\
& \\
 \includegraphics[width=0.48\columnwidth]{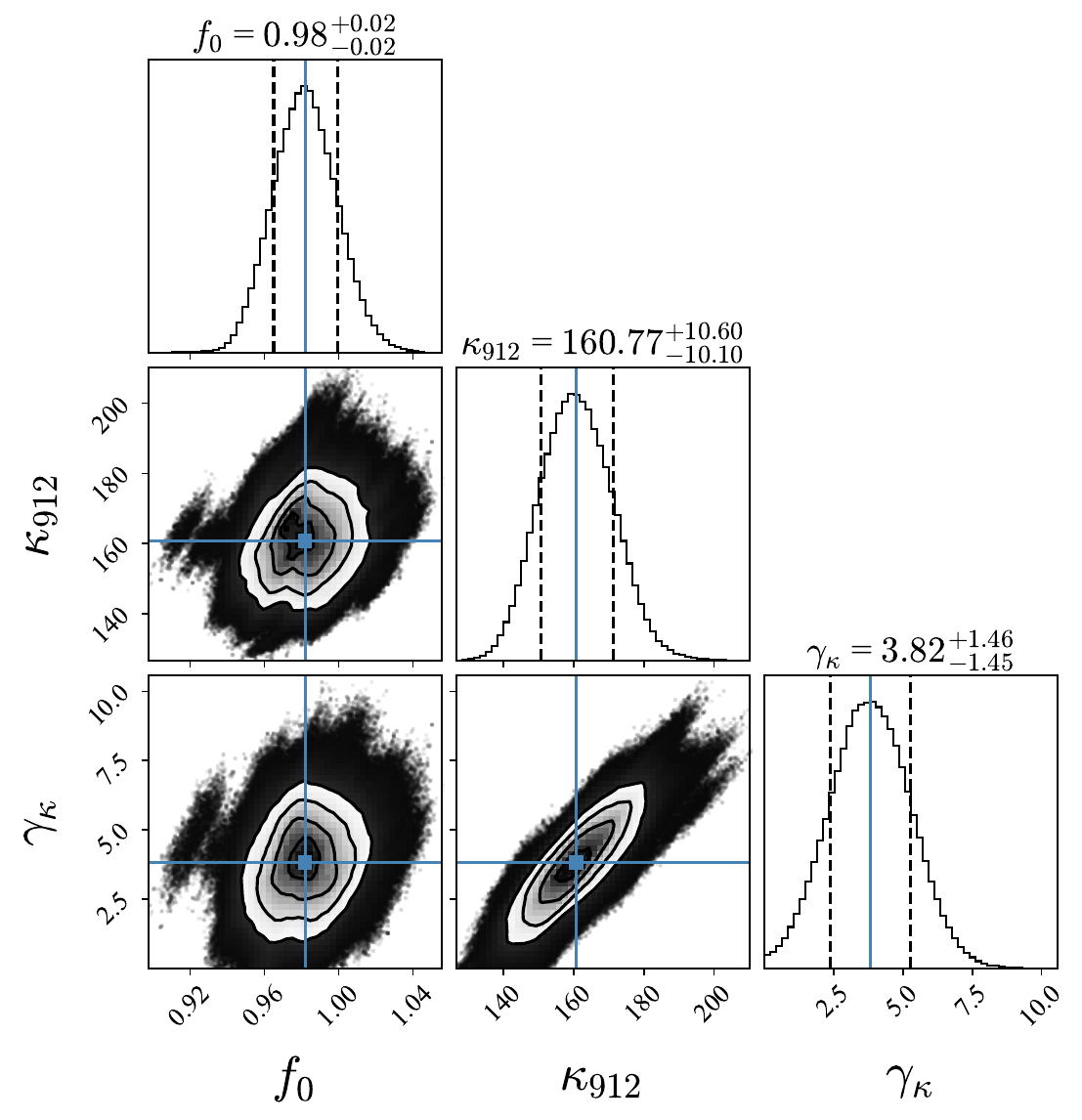} &   \includegraphics[width=0.48\columnwidth]{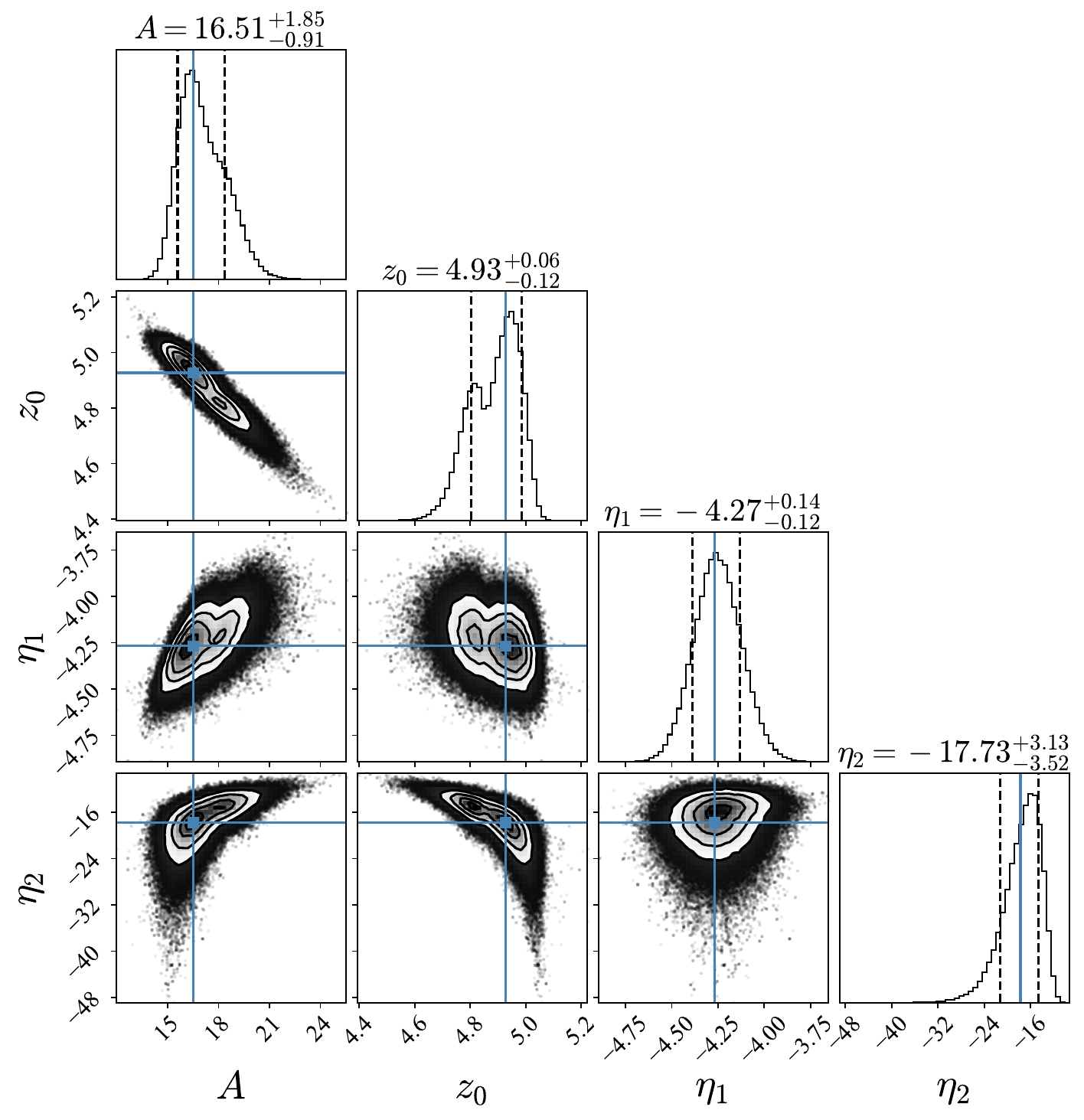} \\
& \\
\end{tabular}

    \caption{{MCMC posteriors. Top left panel: The combined MCMC sampling of the parameter posterior for the $z_\mathrm{median}=3.443$ composite spectrum. The blue lines show the median parameter values in marginal distributions that can be used as an appropriate choice for the maximum posterior estimation. The dash vertical lines denote the $16\%$ and $84\%$ quantiles. Top right panel: The mean free path distribution calculated from the parameter posterior sample in the top-left panel, which is mostly Gaussian-shaped. Lower left panel: The parameter posterior at $z_\mathrm{median}=3.443$ when the parameter $\gamma_\kappa$ in \autoref{eq6} is freed. Lower right panel: The broken power law parameter posterior when fitting the data points in \autoref{fig10}.}}
    \label{figMC}
\end{figure*}

{In this section, we plot the corner plots of all the major MCMC sampling we conducted in \autoref{figMC}. The top panels are an example of fitting the composite spectra with our default model. The lower left panel is an example of fitting with $\gamma_\kappa$ in \autoref{eq4} freed. The lower right panel is the sampling of four parameters in our broken power law fitting in \autoref{fig10}.}

\section{Previous estimates of \lmfp}\label{app1}

We compare our results with the previous estimates that use the similar method to ours, listed in \autoref{tab2}. We do not use the results from { \citet{2024ApJ...965..134D}, }\citet{2023MNRAS.525.4093G} and \citet{2021arXiv210812446B}, as their methods are significantly different from ours, which may introduce unknown biases. We also do not use other early works that first estimate $f(N_{\specnotation{H}{I}},z)$ and calculate $\lmfp$ using \autoref{eq14} (e.g. \citealt{2013ApJ...769..146R}), as they suffer from biases such as line blending and absorber clustering \citep{2014MNRAS.438..476P}.

\begin{table}
  \centering
  \caption{$\lambda^{912}_\mathrm{mfp}$ measurements from literature. The unit of $\lambda^{912}_\mathrm{mfp}$ is $h_{70}^{-1}\,\mathrm{Mpc}$.}
    \begin{tabular}{ccccc}
    \toprule
    $z_\mathrm{qso}$  & $N_\mathrm{qso}$   &  $\lambda^{912}_\mathrm{mfp}$  & $\sigma\left(\lambda^{912}_\mathrm{mfp}\right)$ & Reference \\
    \midrule
    2.09 & 48 & 314.5 & 64.9 & \citet{2018ApJ...860...41L} \\
     &  &  &  & (Low-$z$ sample) \\
    2.44 & 46 & 140.7 & 20.2 & \citet{2018ApJ...860...41L} \\
     &  &  &  & (High-$z$ sample) \\
    2.44 & 53 & 235.8 & 40.3 & \citet{2013ApJ...765..137O} \\
    3.00 & 61 & 110.0 & 34 & \citet{2013ApJ...775...78F} \\
    3.73 & 150 & 52.8 & 5.7 & \citetalias{2009ApJ...705L.113P} \\
    3.78 & 150 & 45 & 4.2 & \citetalias{2009ApJ...705L.113P} \\
    3.83 & 150 & 44.3 & 4.8 & \citetalias{2009ApJ...705L.113P} \\
    3.88 & 150 & 46.5 & 4.8 & \citetalias{2009ApJ...705L.113P} \\
    3.96 & 150 & 38.9 & 3.7 & \citetalias{2009ApJ...705L.113P} \\
    4.07 & 150 & 33 & 3.5 & \citetalias{2009ApJ...705L.113P} \\
    4.22 & 150 & 28.1 & 2.9 & \citetalias{2009ApJ...705L.113P} \\
    4.56 & 57 & 22.2 & 2.3 & \citetalias{2014MNRAS.445.1745W} \\
    4.86 & 49 & 15.1 & 1.8 & \citetalias{2014MNRAS.445.1745W} \\
    5.16 & 39 & 10.3 & 1.6 & \citetalias{2014MNRAS.445.1745W} \\
    5.08 & 44 & 9.33 & +2.06/-1.8 & \citet{2023ApJ...955..115Z} \\
    5.31 & 26 & 5.4 & +1.47/-1.40 & \citet{2023ApJ...955..115Z} \\
    5.65 & 9  & 3.31 & +2.74/-1.34 & \citet{2023ApJ...955..115Z} \\
    5.93 & 18 & 0.81 & +0.73/-0.48 & \citet{2023ApJ...955..115Z} \\
    5.1 & 63 & 9.09  & +1.62/-1.28 & \citet{2021MNRAS.508.1853B} \\
    6.0 & 13 & 0.75  & +0.65/-0.45 & \citet{2021MNRAS.508.1853B} \\
    \bottomrule
    \end{tabular}
  \label{tab2}
\end{table}

\section{Broken Power Law Fitting with More Data}\label{app3}
\begin{figure*}
    \centering
    \includegraphics[width=\textwidth]{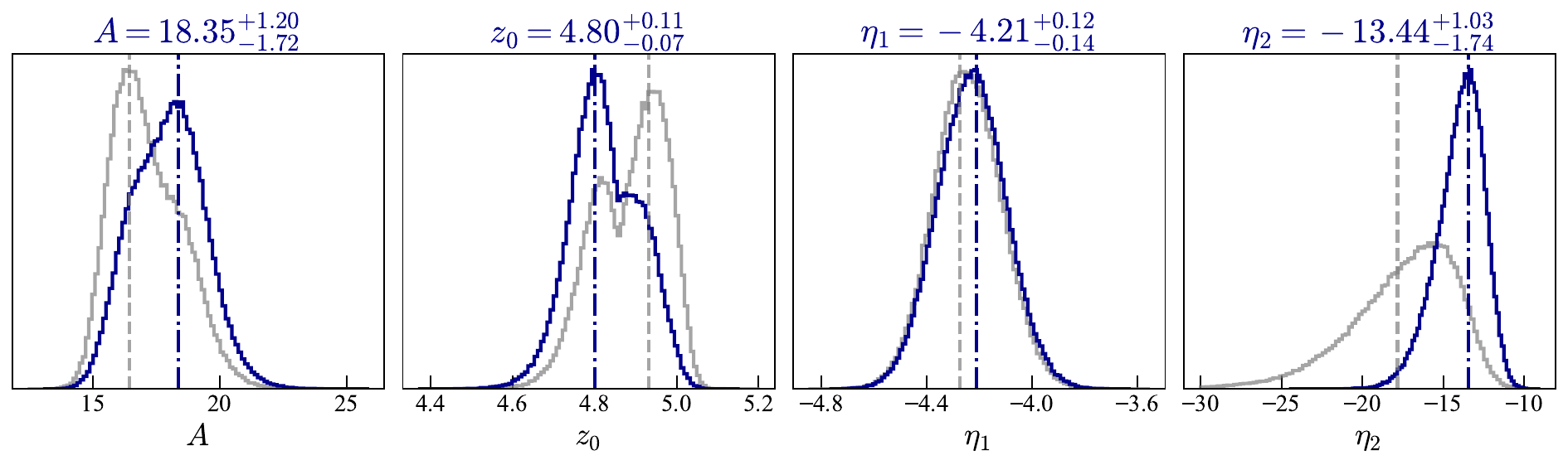}
    \caption{Marginalized parameter posterior of broken power law fitting when incorporating data from \citet{2023MNRAS.525.4093G}, \citet{2024ApJ...965..134D} and \citet{2024MNRAS.533..676S}. The blue histograms are the normalized posterior, the blue vertical dash-dot lines indicate the maximum posterior estimation, and the blue titles are the parameter estimations with the uncertainties defined in the way described in Section~\ref{sec43}. For comparison, we also show the marginalized posterior from Section~\ref{sec43} in grey lines.}
    \label{figapp4}
\end{figure*}

In Section~\ref{sec43} we fit the broken power law model with the measurements in \citetalias{2009ApJ...705L.113P}, \citet{2013ApJ...765..137O}, \citet{2013ApJ...775...78F}, \citetalias{2014MNRAS.445.1745W}, \citet{2018ApJ...860...41L}, \citet{2021MNRAS.508.1853B}, \citet{2023ApJ...955..115Z} and our results. In \citet{2023MNRAS.525.4093G}, \citet{2024ApJ...965..134D} and \citet{2024MNRAS.533..676S}, the authors measured the $\lmfp$ at $5\lesssim z \lesssim 6$ independently with various cosmological radiative transfer simulations. As their methods are mostly model-dependent and thus different from the direct method of others, we do not include their results in our fitting in Section~\ref{sec43}. Here we present the broken power law fitting when incorporating them and show the MCMC parameter posterior in \autoref{figapp4} together with the posterior in Section~\ref{sec43}. The two posteriors are mostly consistent, and biggest difference would be that the broken redshift is lowered to $4.8$. Another interesting thing about the posterior of $z_0$ is that the implicit second peak at $z_0\sim4.8$ in the original posterior is strengthened and becomes the major peak after including more data, implying that the possibility of $z_0=4.80$ cannot be simply excluded and future data are still needed to distinguish the two cases.

\bibliography{ref}
\bibliographystyle{aasjournal}

\end{document}